\documentclass[nofootinbib,aps,prl,twocolumn,showpacs,superscriptaddress]{revtex4-1} 
\usepackage[english]{babel}
\usepackage[T1]{fontenc}
\usepackage[utf8]{inputenc}
\usepackage{amsmath}
\usepackage{graphicx}
\usepackage{amssymb}
\usepackage{multirow}
\usepackage{subfigure}
\usepackage{xspace}
\usepackage{booktabs}

\usepackage[en-US]{datetime2}
\usepackage[left]{lineno}
\usepackage{comment}
\usepackage{soul}
\usepackage{float}
\usepackage{sidecap}
\sidecaptionvpos{figure}{c}
\usepackage{hyperref}

\def\LV{\ifmmode {{\rm LIV}}\else{\scshape LIV}\fi\xspace}
\def\LIV{\ifmmode{{\rm LIV}}\else{\scshape LIV}\fi\xspace}
\def\LI{\ifmmode {{\rm LI}}\else{\scshape LI}\fi\xspace}
\def\Ec{\ifmmode {\mathrm{E_c}}\else{\scshape $ E_c$}\fi\xspace}
\def\hEc{\ifmmode {\mathrm{\hat{E}_c}}\else{\scshape $ \hat{E}_c$}\fi\xspace}
\def\ELIV{\ifmmode {{E_{\LIV}}}\else{\scshape $ E_{\LIV}$}\fi\xspace}
\def\ELIVl{\ifmmode {{E_{\LIV}^{(1)}}}\else{\scshape $ E_{\LIV}^{(1)}$}\fi\xspace}
\def\ELIVn{\ifmmode {{E_{\LIV}^{(n)}}}\else{\scshape $ E_{\LIV}^{(n)}$}\fi\xspace}
\newcommand{\epem}{$e^+e^-$\xspace}

\usepackage[dvipsnames]{xcolor}

\begin{document}

\title{Constraints on Lorentz invariance violation from HAWC observations of gamma rays above 100 TeV}

\author{A.~Albert}
\address{Physics Division, Los Alamos National Laboratory, Los Alamos, NM, USA }

\author{R.~Alfaro}
\address{Instituto de Física, Universidad Nacional Autónoma de México, Ciudad de Mexico, Mexico }

\author{C.~Alvarez}
\address{Universidad Autónoma de Chiapas, Tuxtla Gutiérrez, Chiapas, México}

\author{J.R.~Angeles Camacho}
\address{Instituto de Física, Universidad Nacional Autónoma de México, Ciudad de Mexico, Mexico }

\author{J.C.~Arteaga-Velázquez}
\address{Universidad Michoacana de San Nicolás de Hidalgo, Morelia, Mexico }

\author{K.P.~Arunbabu}
\address{Instituto de Geofísica, Universidad Nacional Autónoma de México, Ciudad de Mexico, Mexico }

\author{D.~Avila Rojas}
\address{Instituto de Física, Universidad Nacional Autónoma de México, Ciudad de Mexico, Mexico }

\author{H.A.~Ayala Solares}
\address{Department of Physics, Pennsylvania State University, University Park, PA, USA }

\author{V.~Baghmanyan}
\address{Institute of Nuclear Physics Polish Academy of Sciences, PL-31342 IFJ-PAN, Krakow, Poland }

\author{E.~Belmont-Moreno}
\address{Instituto de Física, Universidad Nacional Autónoma de México, Ciudad de Mexico, Mexico }

\author{S.Y.~BenZvi}
\address{Department of Physics \& Astronomy, University of Rochester, Rochester, NY , USA }

\author{C.~Brisbois}
\address{Department of Physics, Michigan Technological University, Houghton, MI, USA }

\author{K.S.~Caballero-Mora}
\address{Universidad Autónoma de Chiapas, Tuxtla Gutiérrez, Chiapas, México}

\author{T.~Capistrán}
\address{Instituto Nacional de Astrofísica, Óptica y Electrónica, Puebla, Mexico }

\author{A.~Carramiñana}
\address{Instituto Nacional de Astrofísica, Óptica y Electrónica, Puebla, Mexico }

\author{S.~Casanova}
\address{Institute of Nuclear Physics Polish Academy of Sciences, PL-31342 IFJ-PAN, Krakow, Poland }

\author{U.~Cotti}
\address{Universidad Michoacana de San Nicolás de Hidalgo, Morelia, Mexico }

\author{J.~Cotzomi}
\address{Facultad de Ciencias Físico Matemáticas, Benemérita Universidad Autónoma de Puebla, Puebla, Mexico }

\author{S.~Coutiño de León}
\address{Instituto Nacional de Astrofísica, Óptica y Electrónica, Puebla, Mexico }

\author{E.~De la Fuente}
\address{Departamento de Física, CUCEI, Universidad de Guadalajara, Guadalajara, Mexico} 

\author{C.~de León}
\address{Universidad Michoacana de San Nicolás de Hidalgo, Morelia, Mexico }

\author{B.L.~Dingus}
\address{Physics Division, Los Alamos National Laboratory, Los Alamos, NM, USA }

\author{M.A.~DuVernois}
\address{Department of Physics, University of Wisconsin-Madison, Madison, WI, USA }

\author{J.C.~Díaz-Vélez}
\address{Departamento de Física, Centro Universitario de los Valles, Universidad de Guadalajara, Guadalajara, Mexico}

\author{R.W.~Ellsworth}
\address{Department of Physics, University of Maryland, College Park, MD, USA }

\author{K.~Engel}
\address{Department of Physics, University of Maryland, College Park, MD, USA }

\author{C.~Espinoza}
\address{Instituto de Física, Universidad Nacional Autónoma de México, Ciudad de Mexico, Mexico }

\author{H.~Fleischhack}
\address{Department of Physics, Michigan Technological University, Houghton, MI, USA }

\author{N.~Fraija}
\address{Instituto de Astronomía, Universidad Nacional Autónoma de México, Ciudad de Mexico, Mexico}

\author{A.~Galván-Gámez}
\address{Instituto de Astronomía, Universidad Nacional Autónoma de México, Ciudad de Mexico, Mexico}

\author{D.~Garcia}
\address{Instituto de Física, Universidad Nacional Autónoma de México, Ciudad de Mexico, Mexico }

\author{J.A.~García-González}
\address{Instituto de Física, Universidad Nacional Autónoma de México, Ciudad de Mexico, Mexico }

\author{F.~Garfias}
\address{Instituto de Astronomía, Universidad Nacional Autónoma de México, Ciudad de Mexico, Mexico}

\author{M.M.~González}
\address{Instituto de Astronomía, Universidad Nacional Autónoma de México, Ciudad de Mexico, Mexico}

\author{J.A.~Goodman}
\address{Department of Physics, University of Maryland, College Park, MD, USA }

\author{J.P.~Harding}
\email{jpharding@lanl.gov}
\address{Physics Division, Los Alamos National Laboratory, Los Alamos, NM, USA }

\author{S.~Hernandez}
\address{Instituto de Física, Universidad Nacional Autónoma de México, Ciudad de Mexico, Mexico }

\author{B.~Hona}
\address{Department of Physics, Michigan Technological University, Houghton, MI, USA }

\author{D.~Huang}
\address{Department of Physics, Michigan Technological University, Houghton, MI, USA }

\author{F.~Hueyotl-Zahuantitla}
\address{Universidad Autónoma de Chiapas, Tuxtla Gutiérrez, Chiapas, México}

\author{P.~Hüntemeyer}
\address{Department of Physics, Michigan Technological University, Houghton, MI, USA }

\author{A.~Iriarte}
\address{Instituto de Astronomía, Universidad Nacional Autónoma de México, Ciudad de Mexico, Mexico}

\author{V.~Joshi}
\address{Erlangen Centre for Astroparticle Physics, Friedrich-Alexander-Universit\"{a}t Erlangen-N\"{u}rnberg, Erlangen, Germany}

\author{A.~Lara}
\address{Instituto de Geofísica, Universidad Nacional Autónoma de México, Ciudad de Mexico, Mexico }

\author{W.H.~Lee}
\address{Instituto de Astronomía, Universidad Nacional Autónoma de México, Ciudad de Mexico, Mexico}

\author{H.~León Vargas}
\address{Instituto de Física, Universidad Nacional Autónoma de México, Ciudad de Mexico, Mexico }

\author{J.T.~Linnemann}
\email{linnemann@pa.msu.edu}
\address{Department of Physics and Astronomy, Michigan State University, East Lansing, MI, USA }

\author{A.L.~Longinotti}
\address{Instituto Nacional de Astrofísica, Óptica y Electrónica, Puebla, Mexico }

\author{G.~Luis-Raya}
\address{Universidad Politecnica de Pachuca, Pachuca, Hgo, Mexico }

\author{J.~Lundeen}
\email{lundeenj@msu.edu}
\address{Department of Physics and Astronomy, Michigan State University, East Lansing, MI, USA }

\author{R.~López-Coto}
\address{INFN and Universita di Padova, via Marzolo 8, I-35131,Padova,Italy}

\author{K.~Malone}
\address{Physics Division, Los Alamos National Laboratory, Los Alamos, NM, USA }

\author{S.S.~Marinelli}
\email{samuelmarinelli@gmail.com}
\address{Department of Physics and Astronomy, Michigan State University, East Lansing, MI, USA }

\author{I.~Martinez-Castellanos}
\address{Department of Physics, University of Maryland, College Park, MD, USA }

\author{J.~Martínez-Castro}
\address{Centro de Investigaci\'on en Computaci\'on, Instituto Polit\'ecnico Nacional, M\'exico City, M\'exico.}

\author{H.~Martínez-Huerta}
\email{humbertomh@ifsc.usp.br}
\address{Instituto de F\'isica de S\~ao Carlos, Universidade de S\~ao Paulo, S\~ao Carlos, SP, Brasil}

\author{J.A.~Matthews}
\address{Dept of Physics and Astronomy, University of New Mexico, Albuquerque, NM, USA }

\author{P.~Miranda-Romagnoli}
\address{Universidad Autónoma del Estado de Hidalgo, Pachuca, Mexico }

\author{J.A.~Morales-Soto}
\address{Universidad Michoacana de San Nicolás de Hidalgo, Morelia, Mexico }

\author{E.~Moreno}
\address{Facultad de Ciencias Físico Matemáticas, Benemérita Universidad Autónoma de Puebla, Puebla, Mexico }

\author{A.~Nayerhoda}
\address{Institute of Nuclear Physics Polish Academy of Sciences, PL-31342 IFJ-PAN, Krakow, Poland }

\author{L.~Nellen}
\address{Instituto de Ciencias Nucleares, Universidad Nacional Autónoma de Mexico, Ciudad de Mexico, Mexico }

\author{M.~Newbold}
\address{Department of Physics and Astronomy, University of Utah, Salt Lake City, UT, USA }

\author{M.U.~Nisa}
\address{Department of Physics and Astronomy, Michigan State University, East Lansing, MI, USA }

\author{R.~Noriega-Papaqui}
\address{Universidad Autónoma del Estado de Hidalgo, Pachuca, Mexico }

\author{N.~Omodei}
\affiliation{Stanford University, Stanford, CA 94305, USA}

\author{A.~Peisker}
\address{Department of Physics and Astronomy, Michigan State University, East Lansing, MI, USA }

\author{E.G.~Pérez-Pérez}
\address{Universidad Politecnica de Pachuca, Pachuca, Hgo, Mexico }

\author{C.D.~Rho}
\address{Department of Physics \& Astronomy, University of Rochester, Rochester, NY , USA }

\author{C.~Rivière}
\address{Department of Physics, University of Maryland, College Park, MD, USA }

\author{D.~Rosa-González}
\address{Instituto Nacional de Astrofísica, Óptica y Electrónica, Puebla, Mexico }

\author{M.~Rosenberg}
\address{Department of Physics, Pennsylvania State University, University Park, PA, USA }

\author{E.~Ruiz-Velasco}
\address{Max-Planck Institute for Nuclear Physics, 69117 Heidelberg, Germany}

\author{H.~Salazar}
\address{Facultad de Ciencias Físico Matemáticas, Benemérita Universidad Autónoma de Puebla, Puebla, Mexico }

\author{F.~Salesa Greus}
\address{Institute of Nuclear Physics Polish Academy of Sciences, PL-31342 IFJ-PAN, Krakow, Poland }

\author{A.~Sandoval}
\address{Instituto de Física, Universidad Nacional Autónoma de México, Ciudad de Mexico, Mexico }

\author{M.~Schneider}
\address{Department of Physics, University of Maryland, College Park, MD, USA }

\author{H.~Schoorlemmer}
\address{Max-Planck Institute for Nuclear Physics, 69117 Heidelberg, Germany}

\author{G.~Sinnis}
\address{Physics Division, Los Alamos National Laboratory, Los Alamos, NM, USA }

\author{A.J.~Smith}
\address{Department of Physics, University of Maryland, College Park, MD, USA }

\author{R.W.~Springer}
\address{Department of Physics and Astronomy, University of Utah, Salt Lake City, UT, USA }

\author{P.~Surajbali}
\address{Max-Planck Institute for Nuclear Physics, 69117 Heidelberg, Germany}

\author{E.~Tabachnick}
\address{Department of Physics, University of Maryland, College Park, MD, USA }

\author{M.~Tanner}
\address{Department of Physics, Pennsylvania State University, University Park, PA, USA }

\author{O.~Tibolla}
\address{Universidad Politecnica de Pachuca, Pachuca, Hgo, Mexico }

\author{K.~Tollefson}
\address{Department of Physics and Astronomy, Michigan State University, East Lansing, MI, USA }

\author{I.~Torres}
\address{Instituto Nacional de Astrofísica, Óptica y Electrónica, Puebla, Mexico }

\author{R.~Torres-Escobedo}
\address{Departamento de Física, CUCEI, Universidad de Guadalajara, Guadalajara, Mexico} 
\address{Department of Physics \& Astronomy, Texas Tech University, USA}

\author{T.~Weisgarber}
\address{Department of Physics, University of Wisconsin-Madison, Madison, WI, USA }

\author{G.~Yodh}
\address{Department of Physics and Astronomy, University of California, Irvine, Irvine, CA, USA}

\author{A.~Zepeda}
\address{Physics Department, Centro de Investigacion y de Estudios Avanzados del IPN, Mexico City, DF, Mexico }

\author{H.~Zhou}
\address{Physics Division, Los Alamos National Laboratory, Los Alamos, NM, USA }

\collaboration{HAWC Collaboration}

\date{March, 2020}

\begin{abstract}
Due to the high energies and long distances to the sources, astrophysical observations provide a unique opportunity to test possible signatures of Lorentz invariance violation (LIV).  Superluminal LIV enables the decay of photon at high energy. The High Altitude Water Cherenkov (HAWC) Observatory is among the most sensitive gamma-ray instruments currently operating above 10 TeV. HAWC finds evidence of 100 TeV photon emission from at least four astrophysical sources. These observations exclude, for the strongest of the limits set, the LIV energy scale to $2.2\times10^{31}$ eV, over 1800 times the Planck energy and an improvement of 1 to 2 orders of magnitude over previous limits.
\end{abstract}

\maketitle

\textit{Introduction. -- }
The precise measurements of very high energy~(VHE) photons can be used as a test for fundamental physics, such as the Lorentz symmetry. As for any other fundamental principle, exploring its limits of validity has been an important motivation for theoretical and experimental research. Lorentz invariance (LI) powerfully constrains fundamental interactions of particles and fields. Moreover, theories that go beyond the standard model of particles (SM), such as quantum gravity or string theories, can motivate Lorentz invariance violation (LIV) ~\cite{NAMBU,Bluhm,Pot,ALFARO,QG1,QG2,QG3,QG4,QG5,Gian,SME}. Therefore, the dedicated experimental tests of such effects may also help to clear the path to a unification theory of the fundamental forces of nature. Small LIV effects might occur with unrelated magnitudes in different sectors such as gravitational wave propagation, interactions of gravity and matter, or light propagation.
In the photon sector, some effects of \LIV are expected to increase with energy and over long distances due to cumulative processes in photon propagation. Therefore, astrophysical searches provide sensitive probes of LIV and its potential signatures, such as energy-dependent time delay, photon splitting, vacuum Cherenkov radiation, photon decay, and many other phenomena~\cite{HMH-APL,Hohensee:2008xz,Coleman:1997xq,Klinkhamer:2008ky,Stecker:2003pw,Stecker:2001vb,Vasileiou:2013vra,Astapov:2019xmt,Satunin:2019gsl}.

The High Altitude Water Cherenkov  (HAWC) Observatory is a wide field-of-view array of 300 water tanks, each containing four photomultiplier tube detectors. HAWC is located at 4100 m above sea level at 19º N near the Sierra Negra volcano, in Puebla, Mexico, covering an area of 22,000 m$^2$. Since 2015, HAWC has operated with a live fraction duty cycle greater than 95\%. HAWC recently reported detailed measurements of gamma-ray emission above 100 TeV~\cite{HAWC_CRAB_2019,Abeysekara:2019gov}, made possible thanks to the development of advanced energy reconstruction algorithms, including one using an artificial neural network (NN).

The HAWC observations of high-energy photons in several locations across the sky creates the unique opportunity to test \LIV, through the precise measurement and reconstruction of these VHE photons.
Previous studies of possible LIV constraints with HAWC have indicated its special utility in LIV searches. For instance, Ref.~\cite{HAWC_LIV_GRB} analyzes the possibility to test energy-dependent time delays through GRB and pulsar measurements, which would result in strong limits on \LIV in the photon sector. In~\cite{HAWC_LIV_PD}, the potential of LIV photons to decay to \epem was explored. Further preliminary results were presented in  ~\cite{HAWC_LIV_CPT,HAWC_ICRC19_Humberto}.  

Superluminal LIV allows photon to decay at high energies. Photon decay to light fermions proceeds over short distances (centimeters or less) once above the energy threshold of the process~\cite{HMH-APL,Klinkhamer:2008ky,Hohensee:2008xz,Stecker:2003pw,Stecker:2001vb,Coleman:1997xq}, which would lead to a hard cutoff at high photon energies in astrophysical spectra~\cite{Sam_Thesis}. Another process, photon decay into multiple photons~\cite{Astapov:2019xmt,Satunin:2019gsl,rubstov_MULTI-TEV}, also predicts a significant reduction of the photon flux at VHEs beyond which no photons should reach the Earth from astrophysical distances.

In this work, we study four Galactic sources to determine whether there is a hard cutoff compatible with LIV photon decay in the observed spectra of each source. We find that none of them favor such a phenomenon, and we use recent observations of photons above the energy of 100 TeV with HAWC to improve LIV limits by 1 to 2 orders of magnitude over previous values~\cite{Schreck:2013paa,HMH-APL,Astapov:2019xmt}.
In the next section, we present the highlights of LIV photon decay phenomena. Then, we describe the analysis and present our results, assess systematic uncertainties and sensitivity of our measurements, and finally, present our conclusions.

\textit{Lorentz Invariance Violation -- } 
The introduction of a Lorentz violating term in the SM Lagrangian or spontaneous Lorentz symmetry breaking can induce modifications to the particle dispersion relation, compared to the standard energy-momentum relationship in special relativity~\cite{Coleman:1997xq,Coleman:1998ti,SME}.  Although there are various forms of modified dispersion relation (MDR) for different particles and underlying \LV-theories, several of them lead to similar phenomenology, which can be useful for \LV tests in extreme environments such as the astroparticle scenarios we consider here~\cite{Coleman:1998ti,Coleman:1997xq,GUNTER-PH,GUNTER-PD,HMH-APL,rubstov_MULTI-TEV}. Phenomenologically, the LIV effects can be generalized as a function of energy and momentum. In this way, a family of effective MDRs can be addressed for different particles. The MDR for photons is\footnote{Hereafter, natural units are used, $c=\hbar=1$.},
\begin{equation}\label{eq:GDR}
    E_{\gamma}^{\ 2} -  p_{\gamma}^{\ 2} = \pm |\alpha_{n}|p_\gamma^{\ n+2},
\end{equation}
where $(E_{\gamma},p_{\gamma})$ is the photon four-momentum, $\alpha_{n}$ is the \LV parameter, $n$ is the leading order of the correction from the underlying theory, and 
$p_{\gamma}\approx E_\gamma$ at first order in $\alpha_{n}$~\cite{DIS1,DIS2,DIS3, JACOB, Stecker2009, Stecker2009NJ}. The sign usually refers to the so-called {\it superluminal} ($+$), and {\it subluminal} ($-$) dominant phenomena. For $n>0$, limits on the \LV parameter $\alpha_{n}$  can be interpreted in terms of some LIV energy scale,
\begin{equation}\label{eq:ELIV}
    \ELIVn=\alpha_{n}^{-1/n}\enspace.
\end{equation}
Strong constraints on $\ELIVn$ have been set in astroparticle physics by several techniques~\cite{ GRB-LIV, Schreck:2013paa, VERITAS-LIV,PULSARS-LIV,HESS-LIV, Vasileiou:2013vra, Lang:2018yog, E1, E2, E3}, and below we further constrain it with HAWC observations.

\textit{Photon decays. -- }
Kinematically forbidden processes in classical relativity can be allowed in LIV scenarios, such as vacuum Cherenkov radiation, spontaneous photon emission, photon decay, and photon splitting~\cite{Coleman:1998ti, GUNTER-PH,GUNTER-PD, HMH-APL, rubstov_MULTI-TEV,VCR-17}. The last two could have strong effects on astrophysical photons due to the long distances and the VHE of those processes.  
Here we consider decay into both \epem, and into multiple gamma rays. 

Considering the photon decay, $\gamma \rightarrow e^{+}e^{-}$, due to superluminal LIV, the resulting decay rates are fast and effective at energies where the process is allowed ~\cite{HMH-APL,Proc2,Martinez-Huerta:2017ntv}. This creates a hard cutoff in the gamma-ray spectrum with no high-energy photons reaching the Earth from cosmological distances above a given threshold. The threshold for any order~$n$ is given by
\begin{equation}\label{eq:th}
    \alpha_{n} \le \frac{4 m_e^2}{E_{\gamma}^{\ n}(E_{\gamma}^{\ 2}-4 m_e^{\ 2})},
\end{equation}
where $m_e$ stands for the electron mass~\cite{HMH-APL}. Eqs.~(2) and (3) show that the lower limits on $E_{LIV}^{(n)}$ (upper limits on $\alpha_n$) become more stringent with the increase in the observed photon energy by a factor of $E_{\gamma}^{\ 1+2/n}$ ($E_{\gamma}^{-(n+2)}$ for upper limits on $\alpha_n$).

From Eqs.~(\ref{eq:ELIV}) and~(\ref{eq:th}), we can find \ELIVn for $n=1$ and~$2$,
\begin{equation}\label{eq_limit_1}
    E_{\LIV}^{(1)} \gtrsim 9.57\times 10^{23} {\rm eV} \left(\frac{E_{\gamma}}{\rm TeV} \right)^3, 
\end{equation}
\begin{equation}\label{eq_limit}
    E_{\LIV}^{(2)} \gtrsim 9.78\times 10^{17} {\rm eV} \left(\frac{E_{\gamma}}{\rm TeV} \right)^2. 
\end{equation}
Hence,  a lower limit for $E_{LIV}^{(n)}$ in the photon sector directly emerges from any observed high energy cosmic photon event.
Different fermion decay channels can be explored, but only the lightest $\gamma \rightarrow e^{+}e^{-}$ channel is considered in this paper.
Photon decay in flight from the source leads to a straightforward way to bound LIV that depends primarily on the energy of observed photons, and secondarily on the energy resolution and uncertainties of the detector.

A second superluminal LIV decay process considered in this work is photon splitting to multiple photons,  $\gamma \rightarrow N\gamma$. Refs.~\cite{rubstov_MULTI-TEV,Astapov:2019xmt} show that the dominant splitting process is the photon decay into three photons ($3\gamma$), which has been studied in a model of quantum electrodynamics including LIV and n=2.

The decay rate of photon splitting is~\cite{rubstov_MULTI-TEV,Astapov:2019xmt,Satunin:2019gsl} 
\begin{equation}\label{eq:PS_gamma}
\Gamma_{\gamma \rightarrow 3\gamma} = 5\times 10^{-14} \frac{E_{\gamma}^{\ 19}}{m_e^{\ 8} E_{\LIV}^{(2) \ 10}},
\end{equation}
which is significantly smaller than the photon decay rate considered in the previous section. However, this process has no threshold, and is kinematically allowed whenever $E_\gamma^{\ 2} > p_\gamma^{\ 2}$.
It becomes significant when photons propagate through cosmological distances and also predicts a cutoff at the highest energy part of the photon spectra of astrophysical sources. Despite the lack of a kinematical energy threshold, the strong photon energy dependence of Eq.~(\ref{eq:PS_gamma}) produces an effective one: 
an energy region narrow compared to HAWC’s energy resolution in which the probability for photons to arrive from a source sharply drops.

Because we observe photons from distant sources, we equate the mean free path of a photon to the distance between the source and observer, $L$, that is we take $L~\Gamma~=~1$, with $\Gamma$ translated to units of $kpc^{-1}$. The corresponding LIV limit, as a function of the highest photon energy, is given by,
\begin{equation}\label{eq:PS}
    E_{\LIV}^{(2)} > 
    3.33\times 10^{19} {\rm eV}
    \left(\frac{L}{\rm kpc}\right)^{0.1}\left(\frac{E_\gamma}{\rm TeV}\right)^{1.9}.
\end{equation}
Once again, this photon decay in flight from the source leads to a direct way to bound the LIV energy scale that mainly depends on the highest energy photons observed. It is interesting to note that the higher-order process of  Eq.~(\ref{eq:PS}) produces a stronger limit than the lower order photon decay of Eq.~(\ref{eq_limit}).

Refs.~\cite{rubstov_MULTI-TEV,Astapov:2019xmt,Satunin:2019gsl} discuss a different method of setting limits on subluminal LIV with $n=2$ using modifications to the Bethe-Heitler interaction of photons in the atmosphere. 
However, unlike the photon splitting process, this does not result in a sharp effective threshold. Thus setting a limit using this effect must use different analysis techniques than the ones we have used to analyze the HAWC data, and we must defer such analysis to a later publication.

\textit{Limit Calculation. -- }
Since the emphasis here is on the upper extremes of the spectrum, several details of the HAWC analysis are changed compared to previous analyses such as that of the Crab Nebula spectrum~\cite{HAWC_CRAB_2019}.   
First, we concentrate on the NN energy estimator as it is expected to have better energy resolution ($~.1-.15$ in $\log_{10}$ E/TeV above 50 TeV)~\cite{HAWC_CRAB_2019}. Second, we re-bin the two highest bins of estimated energy, subdividing both the $(100,178)$ and the $(178,316)$ TeV bins into three finer bins each of equal size in log space.

\begin{figure}[t!]
    {\centering
    \includegraphics[width=0.5\textwidth]{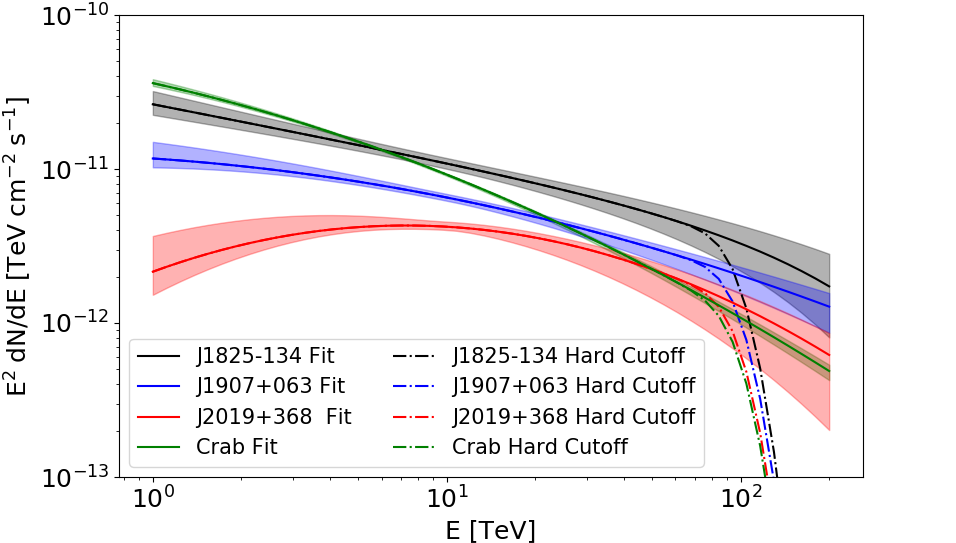}
    }
	\caption{Comparison of the best-fit spectra with those expected were a hard cutoff found at 100 TeV. 
	From top to bottom at 1 TeV: the spectra for the Crab, J1825-134, J1907+063, and J2019+368.  The bands represent statistical uncertainties of the fits.
	}
	\label{fig:spectra}
\end{figure}

We consider the Crab and other three other sources which have evidence of emission above 100 TeV in reconstructed energy~\cite{HAWC_ICRC19_Kelly,Abeysekara:2019gov}. 
For spectral assumptions, we consider a log-parabola for the Crab, eHWC J1907+063, and eHWC J2019+368, and a cutoff-exponential model for eHWC J1825-134, as shown in Fig.~\ref{fig:spectra}.  These choices are consistent with the more detailed information on the sources found in ~\cite{HAWC_CRAB_2019,Abeysekara:2019gov} . 
In analogy with~\cite{Abeysekara:2019gov}, we use the best-fit source position for reconstructed NN energy $>56$ TeV. Finally, to desensitize the results to imperfect modeling of the point spread function, the analysis is carried out in bins with fixed radius about the central position (a so-called top hat bin),
chosen for each source to be large enough that the results no longer depend on the choice of top hat radius; see the Supplemental Material~\cite{Supplemental}.

This analysis provides a lower energy limit, $E_c$, beyond which there is weak or no evidence for the continuation of emission for each source. This lower limit on a hard cutoff also serves as an upper limit on observed photon energy, $E_\gamma$. We perform a fit to the chosen energy spectrum shape and compare the fit likelihood with that of the fit of an energy spectrum convolved with a hard cutoff at energy $E_c$. The hard cutoff is convolved with both the HAWC energy resolution and an additional smoothing of 0.1 in $\log_{10}({\rm E/TeV})$ width to avoid bin edge effects~\cite{Supplemental}. The smoothed hard cutoff is therefore wider than the actual HAWC energy resolution.
Because the hard cutoff model accounts for photons which are mis-reconstructed with energy higher than $E_c$, this test is independent of any assumed spectral shape above $E_c$.
Comparisons of the best-fit spectra with those expected with a hard cutoff at 100 TeV are shown in Fig.~\ref{fig:spectra}.  The source spectra are discussed in detail in~\cite{HAWC_CRAB_2019}.

First, we consider  whether sources show an actual preference for such a hard cutoff. Specifically, we find the profile likelihood (with spectral fit parameters optimized for each $E_c$) as a function of $E_c$ and consider the statistical significance of each value of $\Ec{}$; see~\cite{Supplemental}. The statistical test is to calculate the log-likelihood ratio (details in~\cite{Supplemental}) of the fit with no cutoff and the fit including such a cutoff,  
\begin{equation}
D = 2\ln \left(\frac{\mathcal{L}(\hEc)}{\mathcal{L}(\hEc \rightarrow \infty)}\right).
\end{equation}
where $\hEc$ is the best fit value of $E_c$, and the null hypothesis is the LI limit $\hEc \rightarrow \infty$. We calculate the p-value of observing $D$ or greater
(50\% of D values are 0 since upward fluctuations can\'t drive $\Ec{}$ above $\infty$~\cite{chernoff1954}).  The resulting p-values in the Table~\ref{tabs:results} indicate that none of the sources prefer a cutoff. 
Details of the binned likelihood and treatment of background and forward folding for re\-solution effects are given in~\cite{Supplemental}.
\begin{table}[t!]
\centering
{\begin{tabular}{@{}lccccccc@{}}\toprule
Source & p-value & $E_c(95\%)$ & $E_c(3\sigma)$ \\
\toprule
eHWC J1825-134 & 1.000 & 244 & 158 \\
eHWC J1907+063 & 0.990 & 218 & 162 \\
eHWC J0534+220 (Crab) & 1.000 & 152 & 104 \\
eHWC J2019+368 & 0.828 & 120 & 88 \\
\bottomrule
\end{tabular}
}
\caption{HAWC sources and Photon Energy Limits (TeV).}
\label{tabs:results}
\end{table}

Because our spectra do not indicate a significant preference for $E_c < \infty$, we proceed to set a lower limit on $E_c$, which would occur in LIV photon decay signatures. We consider here two confidence levels (CL): $95\%$ and $99.73\%$ (``$3\sigma$''). The corresponding values of $2\ \Delta \ln \mathcal{L}$ (using Wilks' theorem) for the intervals are $2.71,$ and  $7.74$. These limits are intrinsically one-sided, as we lose statistical power to identify a finite $E_c$ for large values of $E_c$. The results shown in Table~\ref{tabs:results} indicate that we have evidence for greater than 100 TeV emission at \textgreater95\% CL from all four sources and $3\sigma$ evidence from three of them.  More statistical detail can be found in~\cite{Supplemental}.

The 95$\%$ CL limits are reinterpreted as limits on $E_{\gamma}$. Then Eqs.~(\ref{eq_limit_1}), (\ref{eq_limit}), and (\ref{eq:PS}) directly lead to lower limits to $E_{\LIV}^{(1)}$ and $E_{\LIV}^{(2)}$, while  we derive upper limits on $\alpha_0$ from Eq.~(\ref{eq:th}), when $n=0$.
Because a hard photon decay cutoff due to LIV would be at the same energy for any source, we also 
combined the likelihood profiles of all four sources and found an $E_c$ limit of 285 TeV, some $11\%$ higher than the limit from eHWC J1825-134 alone~\cite{Supplemental}. In this way, HAWC can exclude the LIV energy scale of the new physics, $\ELIVl$, to greater than $10^{31}$eV, over 1800 times the Planck energy scale ($\rm E_{Pl}\approx 1.22 \times 10^{28}\ eV$), and more constraining than the best previous values~\cite{HMH-APL,Schreck:2013paa}.
We calculate limits on $E_{\LIV}^{(2)}$ from photon splitting only for individual sources, because the limit depends on the source distance to the observer~\cite{ATNF_Catalog}. These limits are more powerful than the $E_{\LIV}^{(2)}$ limits from photon decay and more constraining than previous values~\cite{Astapov:2019xmt,Satunin:2019gsl}.

\begin{figure}[t!]
    {\raggedleft
    \includegraphics[width=0.5\textwidth]{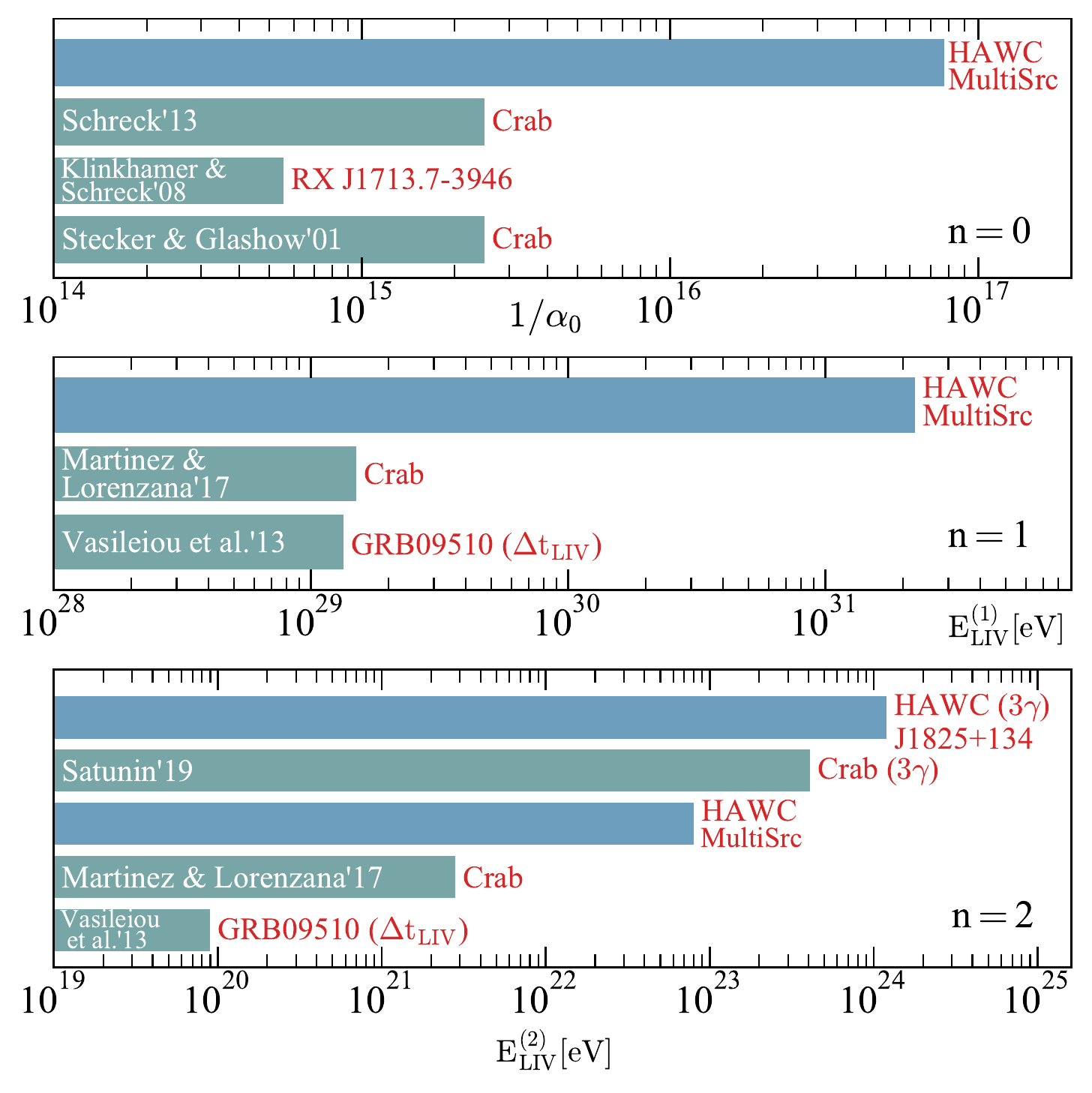}
    }
	\caption{HAWC 95\%~CL LIV limits for $n=0$, 1 and 2. We show previous strong constraints due to photon decay, as well as based on an energy-dependent time delay ($\Delta {\rm t_{LIV}}$) and photon splitting ($3\gamma$). For $n=1$, HAWC limits are orders of magnitude above $\rm E_{Pl}$.  $(\sim10^{28}$~eV~).
	}\label{fig:1}
\end{figure}

We present the HAWC 95\%~CL LIV limits in Table~\ref{tabs:lim}. For comparison, Fig.~\ref{fig:1} shows previous strong limits on photon decay using VHE photons from HEGRA~\cite{HMH-APL,Schreck:2013paa},  
CANGAROO~\cite{Stecker:2001vb}, and HESS~\cite{Klinkhamer:2008ky}. We also show limits due to LIV energy-dependent time delay searches with the {\it Fermi}-LAT~\cite{Vasileiou:2013vra}, and limits due to photon splitting~\cite{Astapov:2019xmt,Satunin:2019gsl}. For a more comprehensive list of these limits and those presented in this work including corresponding values of $\alpha_n$, see the Supplemental Material~\cite{Supplemental}.

We derived the limits above for the LIV coefficients within the general MDR framework, although related limits can also be evaluated in the framework of the Standard Model Extension (SME)~\cite{SME,kos09}. The SME provides a general field-theoretic framework that considers all observer-scalar operators, which are products of the SM and LIV coefficients. The SME coefficients are in general nonisotropic tensors, but their isotropic parts can be written in terms of the corresponding MDR coefficients as described in the Supplemental Material (which also gives constraints from this work on directionally dependent SME coefficients). 
In the SME scenario for n=1 (or any $odd$ $n$), photon decay in SME occurs only for one of the two possible photon polarizations, which involves a drop in photon flux by a factor of 2, see~\cite{Supplemental} for a further discussion. In addition, $odd$ $n$ implies also the effect of birefringence which has been strongly constrained in the SME~\cite{SME_tables}, over 10 orders of magnitude stronger than the constraints to photon decay by the $E_{\LIV}^{(1)}$ excluded here.

\begin{table}
\small \centering
\begin{tabular}{@{}lcccccccc@{}} \toprule
\multicolumn{1}{c}{Source} & 
\begin{tabular}[c]{@{}c@{}}$\rm E_c $     \\ TeV \end{tabular}   & 
\begin{tabular}[c]{@{}c@{}}$\rm L $     \\ kpc \end{tabular}   & 
\begin{tabular}[c]{@{}c@{}}$\alpha_0$     \\ $10^{-17}$\end{tabular}  & 
\begin{tabular}[c]{@{}c@{}}$\rm E_{\LV}^{(1)}$  \\    $10^{31}$eV\end{tabular}  & 
\begin{tabular}[c]{@{}c@{}}$\rm E_{\LV}^{(2)}$  \\    $10^{23}$eV\end{tabular}  &
\begin{tabular}[c]{@{}c@{}}$\rm E_{\LV}^{(2)}$ $_{(3\gamma)}$ \\    $10^{23}$eV\end{tabular}  \\
\toprule
{\bf J1825-134}        
& 244  & 1.55    & 1.75    & 1.39  & 0.58  & 12      \\
{\bf J1907+063}        
& 218  & 2.37     & 2.2     & 0.99  & 0.47  & 10.1      \\
{\bf J0534+220} 
& 152  & 2     & 4.52    & 0.34  & 0.23  &  4.99     \\
{\bf J2019+368}        
& 120  & 1.8    & 7.25    & 0.17  & 0.14  &  3.15      \\
\midrule
{\bf Combined}                  
& 285  & -     & 1.29   & 2.22  & 0.8  & -     \\

\bottomrule
\end{tabular}
\caption{HAWC sources and 95$\%$~CL lower limits on $\rm E_c$, LIV coefficients, and the distance to the observer, L. $\alpha_0$ are upper limits while $E_{\LIV}^{(n)}$ are lower limits. Systematic uncertainties are given in the Supplemental Material~\cite{Supplemental}.
}   
\label{tabs:lim}
\end{table}
%

\textit{Sensitivity and Systematic Uncertainties. -- }  We studied the sensitivity of our method by simulating source instances of the HAWC fit spectra with hard cutoffs, and by computing the expected limits of the HAWC best fit spectra without hard cutoffs. 
These are nearby Galactic sources, for which background light absorption~\cite{Angelis,Gould:1966pza,Gould:1967zza,Gould:1967zzb} is negligible for the distances in Table~\ref{tabs:lim}.
Our simulations of hard cutoffs at 50, 100, and 200 TeV in all spectra resulted in combined fits to $E_c$ within $8\%$ or better of the simulated hard cutoff energy.  The expected combined fit limits had a median of 240 TeV, with 2/3 of the results between 213 and 279 TeV ($-11$ to $+16\%$), suggesting a statistical uncertainty of about $15\%$.  The actual limit of 285 TeV is $+15\%$ higher than the expected median, just over 1$\sigma$.

Following ~\cite{HAWC_CRAB_2019,Abeysekara:2019gov}, we considered a number of systematic uncertainties affecting the LIV limits. 
We summarize them in Table~\ref{tabs:syst}, emphasizing the effects on $E_c$ from the combined limit as this is the most powerful.
Varying simulation parameters in analyzing actual data had relatively minor effects on the results ($-4$ to $7\%$), obtained by adding the effects of all simulation parameters in quadrature.  The most important parameters were phototube efficiency, the time structure of calibration pulses vs. real showers, and charge resolution~\cite{HAWC_CRAB_2019}. 
Using the best spectrum model (best log likelihood among log parabola or powerlaw with exponential cutoff) produced results within $1\%$ of using the 2nd-best spectral shape for all sources.
We also considered the effects of applying a different central source position using all energy bins above 1 TeV
instead of above 56 TeV as the center of the top hat fit, and found the effects to be less than $1\%$. 
Finally,~\cite{HAWC_CRAB_2019} estimates the uncertainty of the absolute HAWC energy scale as a $-6\%$ difference from IACT energy scales at lower energies of $1-30$ TeV. 
Combining these in quadrature gives systematic uncertainty on $\Ec$ of 7$\%$.

The systematic errors have been discussed as fractional effects on $E_c$.  Eq.~(\ref{eq:th}) shows that the fractional uncertainty of $\ELIVn$ will be $(1+2/n)$ (of $\alpha_n$ will be $(n+2)$) times the fractional uncertainty of $E_c$.  The resulting uncertainties are give in the detailed results table in~\cite{Supplemental}.

\begin{table}[ht]
\centering
{\begin{tabular}{@{}lc@{}}
\toprule
Simulation & $-4$\% to +7\% \\
Spectrum choice & $-1$\%  \\
Source location & $-1$\%  \\
Energy scale & $-6$\%  \\ 
\midrule 
Overall &  $-7$\%  to + 7\% \\
\bottomrule
\end{tabular}
}
\caption{Effects of systematic errors on $E_c$ for combined sources.}
\label{tabs:syst}
\end{table}


\textit{Conclusion. -- }
The HAWC Observatory measurements of the highest-energy photons can be used to probe fundamental physics such as violation of Lorentz invariance. In this work, we set \LIV limits by searching for LIV photon decays through the study of four sources with significant high energy emission, including the Crab Nebula. We found that none of them favor a spectrum with a hard cutoff and HAWC finds evidence of 100 TeV photon emission at 95\% CL from four astrophysical sources, with $3\sigma$ evidence from three of them. Furthermore, the dedicated search for such a signature in the spectra increases the energy to which the existence of the most energetic photons can be confirmed, which leads to the new and stringent limits on \LIV in Table~\ref{tabs:lim}, showing an improvement over previous limits of 1-2 orders of magnitude.

\bigskip

\begin{acknowledgments}
The authors are grateful to Alan Kosteleck\'y and Ralf Lehnert for helpful discussions. 
We acknowledge the support from: the US National Science Foundation (NSF); the US Department of Energy Office of High-Energy Physics; the Laboratory Directed Research and Development (LDRD) program of Los Alamos National Laboratory; Consejo Nacional de Ciencia y Tecnolog\'{\'i}a (CONACyT), M{\'e}xico, grants 271051, 232656, 260378, 179588, 254964, 258865, 243290, 132197, A1-S-46288, A1-S-22784, c{\'a}tedras 873, 1563, 341, 323, Red HAWC, M{\'e}xico; DGAPA-UNAM grants AG100317, IN111315, IN111716-3, IN111419, IA102019, IN112218; VIEP-BUAP; PIFI 2012, 2013, PROFOCIE 2014, 2015; FAPESP support No. 2015/15897-1 and 2017/03680-3, and the LNCC/MCTI, Brazil; the University of Wisconsin Alumni Research Foundation; the Institute of Geophysics, Planetary Physics, and Signatures at Los Alamos National Laboratory; Polish Science Centre grant DEC-2018/31/B/ST9/01069, DEC-2017/27/B/ST9/02272; Coordinaci{\'o}n de la Investigaci{\'o}n Cient\'{\'i}fica de la Universidad Michoacana; Royal Society - Newton Advanced Fellowship 180385. Thanks to Scott Delay, Luciano D\'{i}az and Eduardo Murrieta for technical support.
\end{acknowledgments}

\medskip

%
\onecolumngrid
\bigskip

\bigskip

\bigskip

\noindent\rule{17.5cm}{0.3pt}
\section{Supplemental Material \\ for \\ Constraints on Lorentz invariance violation from HAWC observations of gamma rays above 100 TeV}
\begin{center} 
The HAWC Collaboration
\end{center}
\noindent\rule{17.5cm}{0.3pt}

\subsection{Details of photon energy limit calculation  }
For completeness, we exhibit the Poisson binned likelihood in terms of the Poisson likelihood $    P(\ n\ |\ \mu)$.
\begin{equation}
{\mathcal{L}(\hEc )} = \prod_k P(\ n_k\ |\ \mu_k(\hEc)\ )
\end{equation}
where k runs over all the bins of a source, and in the combination limits, over all sources.   

The $\mu_k(\Ec{}) = b_k + S_k(E_c)$ are the Poisson expected counts in a bin of reconstructed energy according to the spectral fit.  $\mu_k$ is a sum of a (hadronic) background term and a signal term derived by simulation from the functional form of the fit spectrum.  
In calculating $D$ the numerator and denominator are calculated separately, that is the fit parameters of the SED are found separately for $\hEc{}$ and $\hEc{}=\infty$.

The signal $S_k$ in bin $k$ in terms of reconstructed energy depends on the spectrum fit and the cutoff energy $\hEc$.   For example for a log parabola fit, the spectrum as a function of true energy $E$, modified by the cutoff, can be written as
\begin{equation}
SED(E) = \Phi_0 (E/E_0)^{-\alpha - \beta ln(E/E_0)} \ f(E,\hEc)
\end{equation}
$E_0$  is not a fit parameter, but a constant chosen to reduce the correlation among the  fit parameters.  The cutoff function $f$ is a step function at $E=\hEc$ convolved with a lognormal of width 0.1; when $\hEc = \infty$, $f = 1$.

The value of $S_k$ is derived from the $SED(E)$ by a parameterization of the simulation-derived energy reconstruction matrix described in the Crab paper~\cite{HAWC_CRAB_2019}.
Conceptually, we can write $S_k(\Ec)$ in terms of a resolution matrix $R$ as 
\begin{equation}
    S_k(\Ec) = \sum_j R_{kj} SED_j(\Ec).
\end{equation}
In this way, the spectrum is forward-folded from true energy to reconstructed energy for the binned maximum likelihood fit, including the ``migration'' of events in one bin of true energy to a different bin in reconstructed energy.  These resolution parameterizations $R$ are available only at specific true energy values.  For fitting smooth spectra, this quantization of the matrix is not problematic; however, if we were to use a hard step function for $f$, this produces irrelevant spikes at bin boundaries and at changepoints of the matrix, which interfere with the optimization of $\Ec{}$.   Smoothing $f$ by approximately the separation between parameterization points removes this problem at the cost of some loss of sensitivity in $\Ec$.

Table \ref{tabs:hibins} gives the summed contents of the upper energy bins for the combined sources. The $\Ec$ analysis uses the full combined likelihood, rather than just the summed bin contents.

\begin{table}[h!]
\centering
{\begin{tabular}{@{}lccccccc@{}}\toprule
Elo$\rm _{bin}$ & N & b & Z(b)& b+migration & Z \\
\toprule
56.2 &	196	&    69.5 &	12.4& 122.1	& 6.1\\
100	  &  28	&    13.2 &	3.6& 22.8	& 1.1\\
121	&    31	&    10.2 & 5.2&  16.3 & 3.2\\
147 &    20 &   7.1	 & 4.0&   10.9	& 2.5\\
178	 &   11	 &   3.7 & 3.1&	    6.0	  &   1.8\\
215	 &   3	 &   2.1 &	0.6&    3.3	   & -0.1\\
261	 &   4	 &   0.9 &	 2.4&   1.4	  &  1.8\\
\bottomrule
\end{tabular}
}
\caption{Bin contents for combined sources: lower bin edge, number of events, hadronic background, significance above background, background + migration from lower bins, significance above background plus migration.}
\label{tabs:hibins}
\end{table}

The migration into a bin is calculated for each bin assuming an $\Ec{}$ placed at the lower bin boundary, so that only the spectrum below the bin contributes to migration into the bin.  The significance for all bins  100 TeV or above is 19 (background only) or 10 (excess over background + migration from below 100 TeV).

Table \ref{tabs:radius} below shows the fixed radius chosen for the various sources.  For comparison, the HAWC point spread function is typically $0.1$-$0.2$ degrees at high energy~\cite{HAWC_CRAB_2019}.  The Crab is an isolated but strong point source; the other sources are extended, and are in busier regions, leading to a different optimization of the chosen radius.    

\begin{table}[h!]
\centering
{\begin{tabular}{@{}lccccccc@{}}\toprule
Source  & radius  \\
\toprule
eHWC J1825-134 & 0.4  \\
eHWC J1907+063 & 0.9 \\
eHWC J0534+220 (Crab) & 0.6 \\
eHWC J2019+368 & 0.8  \\
\bottomrule
\end{tabular}
}
\caption{Angular bin radius (degrees).}
\label{tabs:radius}
\end{table}

\begin{figure}[h!]
    {\centering
    \includegraphics[width=0.6\textwidth]{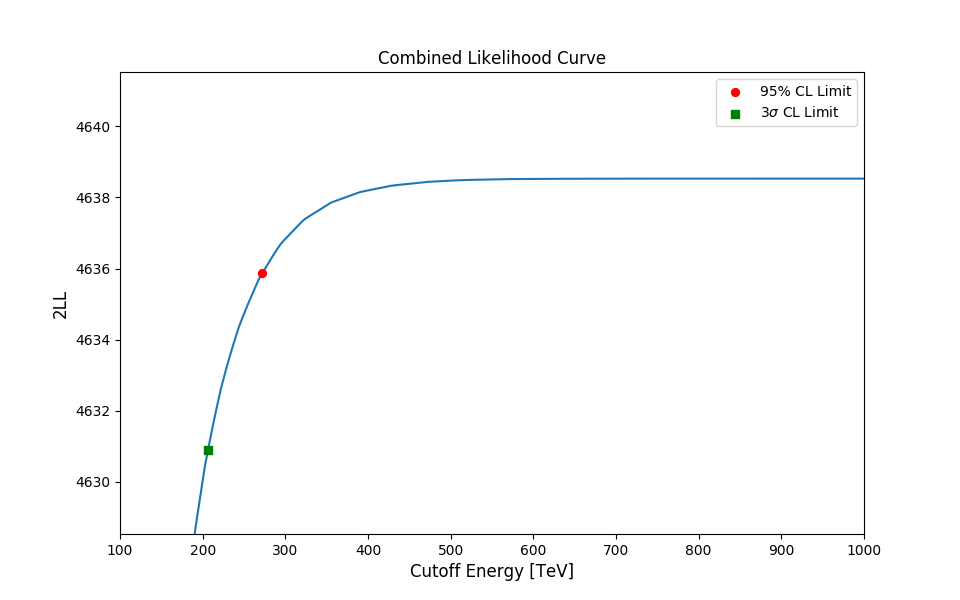}
    }
	\caption{Combined profile likelihood as a function of the Energy cutoff; the upper orange solid point shows the  95\% CL lower limit.}
	\label{fig:combined}
\end{figure}

Fig.~\ref{fig:combined} shows the Combined log-likelihood profile as a function of the energy cutoff. The top and lower points show the lower limits at 95$\%$ CL value and $3\sigma$ CL, respectively. Table I presents the lower limit results for $\rm E_c$  at 95$\%$ for each source and the combined analysis.  Then, by using Eqs.~(\ref{eq:ELIV}),~(\ref{eq:th}), and~(\ref{eq:PS}), we reinterpret the 95$\%$ CL limits as limits on the LIV parameters in Tables \ref{tabs:lim} and \ref{tabs:lim2}.

\subsection{Details of LIV limits }

To aid comparison, Table~\ref{tabs:lim2} details previous strong limits on the decay of very-high energy photons from the HEGRA~\cite{HMH-APL,Schreck:2013paa}, the Tevatron~\cite{Hohensee:2008xz}, CANGAROO~\cite{Stecker:2001vb}, Themistocle~\cite{Coleman:1997xq}, and HESS~\cite{Klinkhamer:2008ky}.
We also show limits due to LIV energy-dependent time delay searches with the {\it Fermi}-LAT~\cite{Vasileiou:2013vra}, 
as well as the limits due to  superluminal photon splitting~\cite{Astapov:2019xmt,Satunin:2019gsl}.

The source distances L in Tables \ref{tabs:lim} and \ref{tabs:lim2} used in calculating the limits on ($3\gamma$) are based on reference~\cite{ATNF_Catalog}.  In the case of eHWC J1825-134, two pulsars lie within the top hat radius used to set the photon energy limit.  The $E^{(2)}_{LIV}$ limit depends on distance as L$^{0.1}$, and we have conservatively chosen the distance to the nearer pulsar, though this makes only a $7\%$ difference in the ($3\gamma$) limit.

\begin{table}[b!]
\small \centering
\begin{tabular}{@{}lccccccccc@{}} \toprule
\multicolumn{1}{c}{Source} & 
\begin{tabular}[c]{@{}c@{}}$\Ec $     \\ TeV \end{tabular}   &
\begin{tabular}[c]{@{}c@{}}$\rm L $     \\ kpc \end{tabular}   & 
\begin{tabular}[c]{@{}c@{}}$\alpha_0$     \\ $10^{-17}$\end{tabular}  & 
\begin{tabular}[c]{@{}c@{}}$\alpha_1$     \\ $10^{-32}$eV$^{-1}$\end{tabular} & 
\begin{tabular}[c]{@{}c@{}}$\alpha_2$     \\ $10^{-48}$eV$^{-2}$\end{tabular} &
\begin{tabular}[c]{@{}c@{}}$\alpha_{2(3\gamma)}$ \\    $10^{-48}$eV$^{-2}$\end{tabular} &
\begin{tabular}[c]{@{}c@{}}$ E_{\rm \LV}^{(1)}$  \\    $10^{31}$eV\end{tabular}  & 
\begin{tabular}[c]{@{}c@{}}$ E_{\rm \LV}^{(2)}$  \\    $10^{23}$eV\end{tabular}  &
\begin{tabular}[c]{@{}c@{}}$ E_{\rm \LV}^{(2)}$ $_{(3\gamma)}$ \\    $10^{23}$eV\end{tabular}  \\
\toprule
{\bf eHWC J1825-134}        
& 244      & 1.55 & $1.75_{1.50}^{2.00}$  & $7.19_{5.68}^{8.70}$  & $295_{212}^{378}$  & $0.70_{0.5}^{0.9}$  & $1.39_{1.10}^{1.68}$  & $0.58_{0.50}^{0.66}$  & $12_{10.3}^{14.0}$      \\
{\bf eHWC J1907+063}        
& 218       & 2.37 & $2.2_{1.89}^{2.51}$   & $10.1_{7.98}^{12.22}$  & $462_{333}^{591}$  & $0.99_{0.71}^{1.27}$  & $0.99_{0.78}^{1.20}$  & $0.47_{0.40}^{0.54}$  & $10.1_{~8.7}^{12.0}$      \\
{\bf eHWC J0534+220 (Crab)} 
& 152       & 2 & $4.52_{3.89}^{5.15}$  & $29.7_{23.46}^{35.94}$  & $1960_{1411}^{2509}$ & $4.01_{2.89}^{5.13}$  & $0.34_{0.27}^{0.41}$  & $0.23_{0.20}^{0.26}$  &  $4.99_{4.3}^{6.0}$     \\
{\bf eHWC J2019+368}        
& 120      & 1.8  & $7.25_{6.24}^{8.27}$  & $60.4_{47.72}^{73.08}$  & $5040_{3629}^{6451}$ & $10.1_{~7.27}^{12.93}$  & $0.17_{0.13}^{0.21}$  & $0.14_{0.12}^{0.16}$  &  $3.15_{2.7}^{4.0}$      \\
\midrule
{\bf Combined}                  
& 285    & -  & $1.29_{1.11}^{1.47}$  & $4.51_{3.56}^{5.46}$   & $158_{114}^{202}$ & -     & $2.22_{1.75}^{2.69}$  & $0.8_{0.69}^{0.91}$  & -     \\
\toprule
Crab (HEGRA) 2017~\cite{HMH-APL}           
& $\sim 56$ &- & -                 & 667  & $1.3\times10^{5}$   & -     & .015  & .028  & -     \\
Tevatron 2016               ~\cite{Hohensee:2008xz}  
& 0.442     &- & $6\times10^{5}$   & -     & -     & -     & -     & -     & -     \\
Crab (HEGRA) 2013 ~\cite{Schreck:2013paa}
& $\sim 56$     &- & 40   & -     & -     & -     & -     & -     & -     \\
RX J1713.7–3946 (HESS) 2008 ~\cite{Klinkhamer:2008ky}  
& 30        &- & 180               & -     & -     & -     & -     & -     & -     \\
Crab (CANGAROO) 2001~\cite{Stecker:2001vb} 	
& 50        &- & 40                & -     & -     & -     & -     & -     & -     \\
Crab (Themistocle) 1997~\cite{Coleman:1997xq}  
& 20        &- & 300               & -     & -     & -     & -     & -     & -     \\
\midrule
{\small GRB09510 ({\it Fermi}-LAT) 2013 $v>c$ ~\cite{Vasileiou:2013vra}}
& - & -    & - & 746  & $1.2\times10^{8}$  & -     & 0.0134 & 0.0009 & -     \\
{\small GRB09510 ({\it Fermi}-LAT) 2013 $v<c$ ~\cite{Vasileiou:2013vra}} 
& - & -   & - & 1075 & $5.9\times10^{7}$   & -     & 0.0093 & 0.0013 & -     \\
\midrule
Crab (Tibet) 2019       ~\cite{Satunin:2019gsl}
& 140 & 2   & - & -     & -         &  5.9   & -     & -     &  4.1   \\
Crab (HEGRA) 2019       ~\cite{Astapov:2019xmt}
& 75 & 2   & - & -     & -         & 59    & -     & -     & 1.3    \\
\bottomrule
\end{tabular}
\caption{HAWC Sources and 95$\%$ CL lower limits on $\Ec$, LIV coefficients, and the distance, L, from the source to the observer. $\alpha_n$ are upper limits while $E_{\rm LIV}^{(n)}$ are lower limits. Subscript $3\gamma$ stands for photon splitting. Errors on the HAWC limits represent the propagation of the systematic errors in Table~\ref{tabs:syst} of the main text. We show previous strong constraints to LIV photon decay (top) as well as the best limits based on an energy-dependent time delay (middle) and superluminal photon splitting (bottom).}   
\label{tabs:lim2}
\end{table}

\subsection{Standard Model Extension limits}

We derived the LIV limits presented in this work in a modified dispersion relation (MDR) framework, although related limits can also be evaluated within the Standard Model Extension (SME)~\cite{SME,kos09}. The SME provides a general field-theoretic framework that considers all observer-scalar operators, which are products of the Standard Model of particle fields and LIV coefficients that may be related to the expectation values of the vectors or tensors of the new physics~\cite{SME}. The new terms can be classified into those that break Charge conjugation, Parity transformation, and Time reversal Symmetry (CPT) and those that preserve CPT. They are called CPT {\it odd} and CPT {\it even}, respectively. In the MDR framework, such classification can be made through the leading order of the correction, when $n$ is {\it odd} or {\it even}. Additionally, the SME can be separated into the sectors of the SM, such as the photon sector, 
\begin{equation}\label{eq:L}
    \mathcal{L}_{photon}^{\rm mSME} = -\frac{1}{4}(k_F)^{\rho \lambda \mu \nu}F_{\rho\lambda}F_{\mu\nu} + (k_{AF})^\mu A^\nu \tilde{F}_{\mu\nu}
\end{equation}
where $A_\mu$ is the gauge field and the field strength tensor is $F_{\mu\nu}=\partial_{\mu}A_ \nu-\partial_{\nu}A_\mu$. $k_F$ and $k_{AF}$ are dimensionless fourth rank tensors and Lorentz-violating coefficients. $k_F$ is dimensionless, while $k_{AF}$ has a mass dimension. By considering only $k_F$, and assuming to have vanishing double trace, to share the symmetries of the Riemann curvature tensor, and restricting the theory to the nonbirefringent and isotropic sector, the number of independent parameters reduces from 256 to 1, $\tilde{\kappa}_{\rm tr}$~\cite{Klinkhamer:2008ky}. In this context, Refs.~\cite{Klinkhamer:2008ky,Schreck:2013paa} report limits to photon decay. The corresponding translation to the MDR coefficient is $\tilde{\kappa}_{\rm tr}  \approx  -\alpha_0/2$ (n=0).

Furthermore, the Lorentz-violating deformation of the photon sector in the SME coefficient can be studied through the decomposition in mass dimension (d) and spherical decomposition ($jm$) of the Lorentz-violating photon dispersion relation from Eq.~(\ref{eq:L})~\cite{kos09}, 
\begin{equation}
    E_{\gamma} \approx \left(1-\zeta^0 \pm \sqrt{(\zeta^1)^2+(\zeta^2)^2+(\zeta^3)^2}\right)p_\gamma,
\end{equation}
where
\begin{equation}
    \zeta^0 = \sum_{d\ jm} p_{\gamma}^{d-4} Y_{jm}(\theta_k,\varphi_k)c^{(d)}_{(I)jm},
\ \ \ \ \ 
    \zeta^3 = \sum_{d\ jm} p_{\gamma}^{d-4} Y_{jm}(\theta_k,\varphi_k)k^{(d)}_{(V)jm}.
\end{equation}

The reinterpretation of the HAWC limits in Tables~ \ref{tabs:lim} and \ref{tabs:lim2}, on these SME coefficients is given as follows.
For $n=2$ (or any $n$ {\it even}), and considering only $\zeta^0$, if there is directional independence (${\small jm = 0~0}$), $c_{(I)~0~0}^{(d=n+4)} = -\sqrt{\pi}~\alpha_{n}$, while in a directional dependent scenario,  $\sum_{jm} Y_{jm}(\theta_k,\varphi_k) c^{(d=n+4)}_{(I)jm} = -\alpha_{n}/2$, where 
$\varphi_{k} =$ right~ascension~(RA)$_{k}$ and  $\theta_{k}=90^{\rm o}-$ declination~(Dec)$_k$, of the source $k$ in a standard Sun-centered inertial reference frame~\cite{Vasileiou:2013vra}.
For $n=1$ (or any $n$ {\it odd}), photon decay in SME occurs only for one of the two possible photon polarizations~\cite{kos09}. A search for this effect involves, instead of a hard cutoff at a threshold, a drop in photon flux by a factor of two at a hard threshold.  This is naturally more difficult to search for, and the corresponding 95\% CL $E_c$ from eHWC J1825-134 drops from 244 to 30 TeV. Considering only $\zeta^3$ and directional independence ($jm=0~0$), the corresponding translation to the MDR coefficient is $k^5_{(V)00} = \sqrt{\pi}~\alpha_1$. In the SME, odd $n$ also implies the effect of birefringence which has been strongly constrained~\cite{SME_tables}, over 10 orders of magnitude stronger than this HAWC limit to photon decay by $k^5_{(V)00}$, even using the 285 TeV combined result.
 
We give the corresponding HAWC limits on SME coefficients, $\tilde{\kappa}_{\rm tr}$, $k^{(5)}_{(V)00}$, $-c_{(I)00}^{(d)}$, and ($-\sum_{jm} Y_{jm}(\theta_k,\varphi_k) c^{(d)}_{(I)jm}$), in the Table~\ref{tabs:SME}.
\begin{table*}[h!]
\small \centering
\begin{tabular}{@{}l|c|c|c|ccc|ccccc@{}} \toprule
 &  $\Ec $ & -$\tilde{\kappa}_{tr}$ & $k^{(5)}_{(V)00}$ &
\multicolumn{3}{c|}{$-c_{(I)00}^{(d)}$} &
\multicolumn{5}{c}{$-\sum_{jm} Y_{jm}(\theta_k,\varphi_k) c^{(d)}_{(I)jm}$}
\\ \multicolumn{1}{c|}{Source} & & & & $\rm d=4$ & $\rm d=6$ & $\rm d=6$~{\small $(3\gamma)$} & 
$\theta_k$ & $\varphi_k$ & $\rm d=4$ & $\rm d=6$ & $\rm d=6$~{\small $(3\gamma)$}
\\ & eV & $10^{-18}$ &$10^{-29}$eV$^{-1}$& $10^{-17}$eV$^{-1}$ & $10^{-49}$eV$^{-2}$ & $10^{-49}$eV$^{-2}$ & º & º 
& $10^{-18}$eV$^{-1}$ & $10^{-49}$eV$^{-2}$ & $10^{-49}$eV$^{-2}$
\\ \toprule
eHWC J1825-134       
& 244       & 8.77  & 6.86 & 3.11   & 5220  &  12.4 
& 103.45  & 276.41    &  8.77  & 1470  & 3.5  \\
eHWC J1907+063        
& 218       & 11  &-& 3.9   & 8200  &   17.5
& 83.75  & 286.95    & 11.5  & 2310  & 4.93  \\
J0534+220 (Crab) 
& 152       & 22.6  &-&  8.01 & 34700  &   71.2
& 67.96  & 83.6    & 22.6   & 9780  & 20.1  \\
eHWC J2019+368        
& 120       & 36.3 &-& 12.9  &  89300 &   178
& 53.26  & 304.94    &  36.3  & 25200  & 50.3  \\ 
\midrule 
{\bf Combined}                    
& 285       & 6.43  &-& 2.28  & 2810  &   -
& -  & - & - & - & -  \\
\bottomrule
\end{tabular}
\caption{HAWC Sources and 95$\%$ CL lower limits on $\rm E_c$ and two-sided LIV limits in the framework of the SME. $(3\gamma)$ stands for the limits derived due to photon splitting. 
}\label{tabs:SME}
\end{table*}


\begin{thebibliography}{59}%
\makeatletter
\providecommand \@ifxundefined [1]{%
 \@ifx{#1\undefined}
}%
\providecommand \@ifnum [1]{%
 \ifnum #1\expandafter \@firstoftwo
 \else \expandafter \@secondoftwo
 \fi
}%
\providecommand \@ifx [1]{%
 \ifx #1\expandafter \@firstoftwo
 \else \expandafter \@secondoftwo
 \fi
}%
\providecommand \natexlab [1]{#1}%
\providecommand \enquote  [1]{``#1''}%
\providecommand \bibnamefont  [1]{#1}%
\providecommand \bibfnamefont [1]{#1}%
\providecommand \citenamefont [1]{#1}%
\providecommand \href@noop [0]{\@secondoftwo}%
\providecommand \href [0]{\begingroup \@sanitize@url \@href}%
\providecommand \@href[1]{\@@startlink{#1}\@@href}%
\providecommand \@@href[1]{\endgroup#1\@@endlink}%
\providecommand \@sanitize@url [0]{\catcode `\\12\catcode `\$12\catcode
  `\&12\catcode `\#12\catcode `\^12\catcode `\_12\catcode `\%12\relax}%
\providecommand \@@startlink[1]{}%
\providecommand \@@endlink[0]{}%
\providecommand \url  [0]{\begingroup\@sanitize@url \@url }%
\providecommand \@url [1]{\endgroup\@href {#1}{\urlprefix }}%
\providecommand \urlprefix  [0]{URL }%
\providecommand \Eprint [0]{\href }%
\providecommand \doibase [0]{http://dx.doi.org/}%
\providecommand \selectlanguage [0]{\@gobble}%
\providecommand \bibinfo  [0]{\@secondoftwo}%
\providecommand \bibfield  [0]{\@secondoftwo}%
\providecommand \translation [1]{[#1]}%
\providecommand \BibitemOpen [0]{}%
\providecommand \bibitemStop [0]{}%
\providecommand \bibitemNoStop [0]{.\EOS\space}%
\providecommand \EOS [0]{\spacefactor3000\relax}%
\providecommand \BibitemShut  [1]{\csname bibitem#1\endcsname}%
\let\auto@bib@innerbib\@empty
\medskip
\bibitem [{\citenamefont {Nambu}(1968)}]{NAMBU}%
  \BibitemOpen
  \bibfield  {author} {\bibinfo {author} {\bibfnamefont {Y.}~\bibnamefont
  {Nambu}},\ }\href@noop {} {\bibfield  {journal} {\bibinfo  {journal}
  {Supp. Prog. Theor. Phys.}\ }\textbf {\bibinfo
  {volume} {E68}},\ \bibinfo {pages} {190} (\bibinfo {year}
  {1968})}\BibitemShut {NoStop}%
\bibitem [{\citenamefont {Bluhm}(2014)}]{Bluhm}%
  \BibitemOpen
  \bibfield  {author} {\bibinfo {author} {\bibfnamefont {R.}~\bibnamefont
  {Bluhm}},\ }in\ \href {\doibase 10.1007/978-3-642-41992-8_23} {\emph
  {\bibinfo {booktitle} {Springer Handbook of Spacetime}}},\ \bibinfo {editor}
  {edited by\ \bibinfo {editor} {\bibfnamefont {A.}~\bibnamefont {Ashtekar}}\
  and\ \bibinfo {editor} {\bibfnamefont {V.}~\bibnamefont {Petkov}}}\ (\bibinfo
  {year} {2014})\ pp.\ \bibinfo {pages} {485--507},\ \Eprint
  {http://arxiv.org/abs/1302.1150} {arXiv:1302.1150 [hep-ph]} \BibitemShut
  {NoStop}%
\bibitem [{\citenamefont {Potting}(2013)}]{Pot}%
  \BibitemOpen
  \bibfield  {author} {\bibinfo {author} {\bibfnamefont {R.}~\bibnamefont
  {Potting}},\ } \href
  {\doibase 10.1088/1742-6596/447/1/012009} {\bibfield  {journal} {\bibinfo
  {journal} {J. Phys. Conf. Ser.}\ }\textbf {\bibinfo {volume} {447}},\
  \bibinfo {pages} {012009} (\bibinfo {year} {2013})}\BibitemShut {NoStop}%
\bibitem [{\citenamefont {Alfaro}(2005)}]{ALFARO}%
  \BibitemOpen
  \bibfield  {author} {\bibinfo {author} {\bibfnamefont {J.}~\bibnamefont
  {Alfaro}},\ }\href {\doibase 10.1103/PhysRevLett.94.221302} {\bibfield
  {journal} {\bibinfo  {journal} {Phys. Rev. Lett.}\ }\textbf {\bibinfo
  {volume} {94}},\ \bibinfo {pages} {221302} (\bibinfo {year} {2005})},\
  \Eprint {http://arxiv.org/abs/hep-th/0412295} {arXiv:hep-th/0412295 [hep-th]}
  \BibitemShut {NoStop}%
\bibitem [{\citenamefont {Amelino-Camelia}(2001{\natexlab{a}})}]{QG1}%
  \BibitemOpen
  \bibfield  {author} {\bibinfo {author} {\bibfnamefont {G.}~\bibnamefont
  {Amelino-Camelia}},\ }\href {\doibase 10.1038/35074035} {\bibfield  {journal}
  {\bibinfo  {journal} {Nature}\ }\textbf {\bibinfo {volume} {410}},\ \bibinfo
  {pages} {1065} (\bibinfo {year} {2001}{\natexlab{a}})},\ \Eprint
  {http://arxiv.org/abs/gr-qc/0104086} {arXiv:gr-qc/0104086 [gr-qc]}
  \BibitemShut {NoStop}%
\bibitem [{\citenamefont {Ellis}\ \emph
  {et~al.}(2000{\natexlab{a}})\citenamefont {Ellis}, \citenamefont
  {Mavromatos},\ and\ \citenamefont {Nanopoulos}}]{QG2}%
  \BibitemOpen
  \bibfield  {author} {\bibinfo {author} {\bibfnamefont {J.~R.}\ \bibnamefont
  {Ellis}}, \bibinfo {author} {\bibfnamefont {N.~E.}\ \bibnamefont
  {Mavromatos}}, \ and\ \bibinfo {author} {\bibfnamefont {D.~V.}\ \bibnamefont
  {Nanopoulos}},\ }\href {\doibase 10.1023/A:1001852601248} {\bibfield
  {journal} {\bibinfo  {journal} {Gen. Relativ. Gravit.}\ }\textbf {\bibinfo {volume}
  {32}},\ \bibinfo {pages} {127} (\bibinfo {year} {2000}{\natexlab{a}})},\
  \Eprint {http://arxiv.org/abs/gr-qc/9904068} {arXiv:gr-qc/9904068 [gr-qc]}
  \BibitemShut {NoStop}%
\bibitem [{\citenamefont {Ellis}\ \emph
  {et~al.}(2000{\natexlab{b}})\citenamefont {Ellis}, \citenamefont
  {Mavromatos}, \citenamefont {Nanopoulos},\ and\ \citenamefont
  {Volkov}}]{QG3}%
  \BibitemOpen
  \bibfield  {author} {\bibinfo {author} {\bibfnamefont {J.~R.}\ \bibnamefont
  {Ellis}}, \bibinfo {author} {\bibfnamefont {N.~E.}\ \bibnamefont
  {Mavromatos}}, \bibinfo {author} {\bibfnamefont {D.~V.}\ \bibnamefont
  {Nanopoulos}}, \ and\ \bibinfo {author} {\bibfnamefont {G.}~\bibnamefont
  {Volkov}},\ }\href {\doibase 10.1023/A:1001980530113} {\bibfield  {journal}
  {\bibinfo  {journal} {Gen. Relativ. Gravit.}\ }\textbf {\bibinfo {volume} {32}},\
  \bibinfo {pages} {1777} (\bibinfo {year} {2000}{\natexlab{b}})},\ \Eprint
  {http://arxiv.org/abs/gr-qc/9911055} {arXiv:gr-qc/9911055 [gr-qc]}
  \BibitemShut {NoStop}%
\bibitem [{\citenamefont {Ellis}\ \emph {et~al.}(1999)\citenamefont {Ellis},
  \citenamefont {Mavromatos},\ and\ \citenamefont {Nanopoulos}}]{QG4}%
  \BibitemOpen
  \bibfield  {author} {\bibinfo {author} {\bibfnamefont {J.~R.}\ \bibnamefont
  {Ellis}}, \bibinfo {author} {\bibfnamefont {N.~E.}\ \bibnamefont
  {Mavromatos}}, \ and\ \bibinfo {author} {\bibfnamefont {D.~V.}\ \bibnamefont
  {Nanopoulos}},\ }\href {\doibase 10.1103/PhysRevD.61.027503} {\bibfield
  {journal} {\bibinfo  {journal} {Phys. Rev.}\ }\textbf {\bibinfo {volume}
  {D61}},\ \bibinfo {pages} {027503} (\bibinfo {year} {1999})},\ \Eprint
  {http://arxiv.org/abs/gr-qc/9906029} {arXiv:gr-qc/9906029 [gr-qc]}
  \BibitemShut {NoStop}%
\bibitem [{\citenamefont {Gambini}\ and\ \citenamefont {Pullin}(1999)}]{QG5}%
  \BibitemOpen
  \bibfield  {author} {\bibinfo {author} {\bibfnamefont {R.}~\bibnamefont
  {Gambini}}\ and\ \bibinfo {author} {\bibfnamefont {J.}~\bibnamefont
  {Pullin}},\ }\href {\doibase 10.1103/PhysRevD.59.124021} {\bibfield
  {journal} {\bibinfo  {journal} {Phys. Rev.}\ }\textbf {\bibinfo {volume}
  {D59}},\ \bibinfo {pages} {124021} (\bibinfo {year} {1999})},\ \Eprint
  {http://arxiv.org/abs/gr-qc/9809038} {arXiv:gr-qc/9809038 [gr-qc]}
  \BibitemShut {NoStop}%
\bibitem [{\citenamefont {Calcagni}(2017)}]{Gian}%
  \BibitemOpen
  \bibfield  {author} {\bibinfo {author} {\bibfnamefont {G.}~\bibnamefont
  {Calcagni}},\ }\href {\doibase 10.1140/epjc/s10052-017-4841-6} {\bibfield
  {journal} {\bibinfo  {journal} {Eur. Phys. J.}\ }\textbf {\bibinfo {volume}
  {C77}},\ \bibinfo {pages} {291} (\bibinfo {year} {2017})},\ \Eprint
  {http://arxiv.org/abs/1603.03046} {arXiv:1603.03046 [gr-qc]} \BibitemShut
  {NoStop}%
\bibitem [{\citenamefont {Colladay}\ and\ \citenamefont
  {Kosteleck\'y}(1998)}]{SME}%
  \BibitemOpen
  \bibfield  {author} {\bibinfo {author} {\bibfnamefont {D.}~\bibnamefont
  {Colladay}}\ and\ \bibinfo {author} {\bibfnamefont {V.~A.}\ \bibnamefont
  {Kosteleck\'y}},\ }\href {\doibase 10.1103/PhysRevD.58.116002} {\bibfield
  {journal} {\bibinfo  {journal} {Phys. Rev.}\ }\textbf {\bibinfo {volume}
  {D58}},\ \bibinfo {pages} {116002} (\bibinfo {year} {1998})},\ \Eprint
  {http://arxiv.org/abs/hep-ph/9809521} {arXiv:hep-ph/9809521 [hep-ph]}
  \BibitemShut {NoStop}%
\bibitem [{\citenamefont {Mart\'inez-Huerta}\ and\ \citenamefont
  {P\'erez-Lorenzana}(2017)}]{HMH-APL}%
  \BibitemOpen
  \bibfield  {author} {\bibinfo {author} {\bibfnamefont {H.}~\bibnamefont
  {Mart\'inez-Huerta}}\ and\ \bibinfo {author} {\bibfnamefont {A.}~\bibnamefont
  {P\'erez-Lorenzana}},\ }\href {\doibase 10.1103/PhysRevD.95.063001}
  {\bibfield  {journal} {\bibinfo  {journal} {Phys. Rev.}\ }\textbf {\bibinfo
  {volume} {D95}},\ \bibinfo {pages} {063001} (\bibinfo {year} {2017})},\
  \Eprint {http://arxiv.org/abs/1610.00047} {arXiv:1610.00047 [astro-ph.HE]}
  \BibitemShut {NoStop}%
\bibitem [{\citenamefont {Hohensee}\ \emph {et~al.}(2009)\citenamefont
  {Hohensee}, \citenamefont {Lehnert}, \citenamefont {Phillips},\ and\
  \citenamefont {Walsworth}}]{Hohensee:2008xz}%
  \BibitemOpen
  \bibfield  {author} {\bibinfo {author} {\bibfnamefont {M.~A.}\ \bibnamefont
  {Hohensee}}, \bibinfo {author} {\bibfnamefont {R.}~\bibnamefont {Lehnert}},
  \bibinfo {author} {\bibfnamefont {D.~F.}\ \bibnamefont {Phillips}}, \ and\
  \bibinfo {author} {\bibfnamefont {R.~L.}\ \bibnamefont {Walsworth}},\ }\href
  {\doibase 10.1103/PhysRevD.80.036010} {\bibfield  {journal} {\bibinfo
  {journal} {Phys. Rev.}\ }\textbf {\bibinfo {volume} {D80}},\ \bibinfo {pages}
  {036010} (\bibinfo {year} {2009})},\ \Eprint {http://arxiv.org/abs/0809.3442}
  {arXiv:0809.3442 [hep-ph]} \BibitemShut {NoStop}%
\bibitem [{\citenamefont {Coleman}\ and\ \citenamefont
  {Glashow}(1997)}]{Coleman:1997xq}%
  \BibitemOpen
  \bibfield  {author} {\bibinfo {author} {\bibfnamefont {S.~R.}\ \bibnamefont
  {Coleman}}\ and\ \bibinfo {author} {\bibfnamefont {S.~L.}\ \bibnamefont
  {Glashow}},\ }\href {\doibase 10.1016/S0370-2693(97)00638-2} {\bibfield
  {journal} {\bibinfo  {journal} {Phys. Lett.}\ }\textbf {\bibinfo {volume}
  {B405}},\ \bibinfo {pages} {249} (\bibinfo {year} {1997})},\ \Eprint
  {http://arxiv.org/abs/hep-ph/9703240} {arXiv:hep-ph/9703240 [hep-ph]}
  \BibitemShut {NoStop}%
\bibitem [{\citenamefont {Klinkhamer}\ and\ \citenamefont
  {Schreck}(2008)}]{Klinkhamer:2008ky}%
  \BibitemOpen
  \bibfield  {author} {\bibinfo {author} {\bibfnamefont {F.~R.}\ \bibnamefont
  {Klinkhamer}}\ and\ \bibinfo {author} {\bibfnamefont {M.}~\bibnamefont
  {Schreck}},\ }\href {\doibase 10.1103/PhysRevD.78.085026} {\bibfield
  {journal} {\bibinfo  {journal} {Phys. Rev.}\ }\textbf {\bibinfo {volume}
  {D78}},\ \bibinfo {pages} {085026} (\bibinfo {year} {2008})},\ \Eprint
  {http://arxiv.org/abs/0809.3217} {arXiv:0809.3217 [hep-ph]} \BibitemShut
  {NoStop}%
\bibitem [{\citenamefont {Stecker}(2003)}]{Stecker:2003pw}%
  \BibitemOpen
  \bibfield  {author} {\bibinfo {author} {\bibfnamefont {F.~W.}\ \bibnamefont
  {Stecker}},\ }\href {\doibase 10.1016/j.astropartphys.2003.08.006} {\bibfield
   {journal} {\bibinfo  {journal} {Astropart. Phys.}\ }\textbf {\bibinfo
  {volume} {20}},\ \bibinfo {pages} {85} (\bibinfo {year} {2003})},\ \Eprint
  {http://arxiv.org/abs/astro-ph/0308214} {arXiv:astro-ph/0308214 [astro-ph]}
  \BibitemShut {NoStop}%
\bibitem [{\citenamefont {Stecker}\ and\ \citenamefont
  {Glashow}(2001)}]{Stecker:2001vb}%
  \BibitemOpen
  \bibfield  {author} {\bibinfo {author} {\bibfnamefont {F.~W.}\ \bibnamefont
  {Stecker}}\ and\ \bibinfo {author} {\bibfnamefont {S.~L.}\ \bibnamefont
  {Glashow}},\ }\href {\doibase 10.1016/S0927-6505(01)00137-2} {\bibfield
  {journal} {\bibinfo  {journal} {Astropart. Phys.}\ }\textbf {\bibinfo
  {volume} {16}},\ \bibinfo {pages} {97} (\bibinfo {year} {2001})},\ \Eprint
  {http://arxiv.org/abs/astro-ph/0102226} {arXiv:astro-ph/0102226 [astro-ph]}
  \BibitemShut {NoStop}%
\bibitem [{\citenamefont {Vasileiou}\ \emph {et~al.}(2013)\citenamefont
  {Vasileiou}, \citenamefont {Jacholkowska}, \citenamefont {Piron},
  \citenamefont {Bolmont}, \citenamefont {Couturier}, \citenamefont {Granot},
  \citenamefont {Stecker}, \citenamefont {Cohen-Tanugi},\ and\ \citenamefont
  {Longo}}]{Vasileiou:2013vra}%
  \BibitemOpen
  \bibfield  {author} {\bibinfo {author} {\bibfnamefont {V.}~\bibnamefont
  {Vasileiou}}, \bibinfo {author} {\bibfnamefont {A.}~\bibnamefont
  {Jacholkowska}}, \bibinfo {author} {\bibfnamefont {F.}~\bibnamefont {Piron}},
  \bibinfo {author} {\bibfnamefont {J.}~\bibnamefont {Bolmont}}, \bibinfo
  {author} {\bibfnamefont {C.}~\bibnamefont {Couturier}}, \bibinfo {author}
  {\bibfnamefont {J.}~\bibnamefont {Granot}}, \bibinfo {author} {\bibfnamefont
  {F.~W.}\ \bibnamefont {Stecker}}, \bibinfo {author} {\bibfnamefont
  {J.}~\bibnamefont {Cohen-Tanugi}}, \ and\ \bibinfo {author} {\bibfnamefont
  {F.}~\bibnamefont {Longo}},\ }\href {\doibase 10.1103/PhysRevD.87.122001}
  {\bibfield  {journal} {\bibinfo  {journal} {Phys. Rev.}\ }\textbf {\bibinfo
  {volume} {D87}},\ \bibinfo {pages} {122001} (\bibinfo {year} {2013})},\
  \Eprint {http://arxiv.org/abs/1305.3463} {arXiv:1305.3463 [astro-ph.HE]}
  \BibitemShut {NoStop}%
\bibitem [{\citenamefont {Astapov}\ \emph {et~al.}(2019)\citenamefont
  {Astapov}, \citenamefont {Kirpichnikov},\ and\ \citenamefont
  {Satunin}}]{Astapov:2019xmt}%
  \BibitemOpen
  \bibfield  {author} {\bibinfo {author} {\bibfnamefont {K.}~\bibnamefont
  {Astapov}}, \bibinfo {author} {\bibfnamefont {D.}~\bibnamefont
  {Kirpichnikov}}, \ and\ \bibinfo {author} {\bibfnamefont {P.}~\bibnamefont
  {Satunin}},\ }\href {\doibase 10.1088/1475-7516/2019/04/054} {\bibfield
  {journal} {\bibinfo  {journal} {JCAP}\ }\textbf {\bibinfo {volume} {1904}},\
  \bibinfo {pages} {054} (\bibinfo {year} {2019})},\ \Eprint
  {http://arxiv.org/abs/1903.08464} {arXiv:1903.08464 [hep-ph]} \BibitemShut
  {NoStop}%
\bibitem [{\citenamefont {Satunin}(2019)}]{Satunin:2019gsl}%
  \BibitemOpen
  \bibfield  {author} {\bibinfo {author} {\bibfnamefont {P.}~\bibnamefont
  {Satunin}},\ }\href {\doibase 10.1140/epjc/s10052-019-7520-y} {\bibfield
  {journal} {\bibinfo  {journal} {Eur. Phys. J.}\ }\textbf {\bibinfo {volume}
  {C79}},\ \bibinfo {pages} {1011} (\bibinfo {year} {2019})},\ \Eprint
  {http://arxiv.org/abs/1906.08221} {arXiv:1906.08221 [astro-ph.HE]}
  \BibitemShut {NoStop}%
\bibitem [{\citenamefont {Abeysekara}\ \emph {et~al.}(2019)\citenamefont
  {Abeysekara} \emph {et~al.}}]{HAWC_CRAB_2019}%
  \BibitemOpen
  \bibfield  {author} {\bibinfo {author} {\bibfnamefont {A.~U.}\ \bibnamefont
  {Abeysekara}} \emph {et~al.} (\bibinfo {collaboration} {HAWC}),\ }\href
  {\doibase 10.3847/1538-4357/ab2f7d} {\bibfield  {journal} {\bibinfo
  {journal} {Astrophys. J.}\ }\textbf {\bibinfo {volume} {881}},\
  \bibinfo {pages} {134} (\bibinfo {year} {2019})},\ \Eprint
  {http://arxiv.org/abs/1905.12518} {arXiv:1905.12518 [astro-ph.HE]}
  \BibitemShut {NoStop}%
\bibitem [{\citenamefont {Abeysekara}\ \emph {et~al.}(2020)\citenamefont
  {Abeysekara} \emph {et~al.}}]{Abeysekara:2019gov}%
  \BibitemOpen
  \bibfield  {author} {\bibinfo {author} {\bibfnamefont {A.~U.}\ \bibnamefont
  {Abeysekara}} \emph {et~al.} (\bibinfo {collaboration} {HAWC}),\ }\href
  {\doibase 10.1103/PhysRevLett.124.021102} {\bibfield  {journal} {\bibinfo
  {journal} {Phys. Rev. Lett.}\ }\textbf {\bibinfo {volume} {124}},\ \bibinfo
  {pages} {021102} (\bibinfo {year} {2020})},\ \Eprint
  {http://arxiv.org/abs/1909.08609} {arXiv:1909.08609 [astro-ph.HE]}
  \BibitemShut {NoStop}%
\bibitem [{\citenamefont {Nellen}(2016)}]{HAWC_LIV_GRB}%
  \BibitemOpen
  \bibfield  {author} {\bibinfo {author} {\bibfnamefont {L.}~\bibnamefont
  {Nellen}} (\bibinfo {collaboration} {HAWC Collaboration}),\ } \href {\doibase 10.22323/1.236.0850} {\bibinfo{journal}{Proc. of Sci. ICRC2015} {\bf \bibinfo{volume}{850}}}\ (\bibinfo {year}
  {2016}),\ \Eprint {http://arxiv.org/abs/1508.03930} {arXiv:1508.03930
  [astro-ph.HE]} \BibitemShut {NoStop}%
\bibitem [{\citenamefont {Martínez-Huerta}(2018)}]{HAWC_LIV_PD}%
  \BibitemOpen
  \bibfield  {author} {\bibinfo {author} {\bibfnamefont {H.}~\bibnamefont
  {Martínez-Huerta}} (\bibinfo {collaboration} {HAWC Collaboration}),\ }\href {\doibase
  10.22323/1.301.0868} {\bibfield  {journal} {\bibinfo  {journal} {Proc. of Sci. ICRC2017}\
  }\textbf {\bibinfo {volume} {868}},\ (\bibinfo
  {year} {2018})},\ \Eprint
  {http://arxiv.org/abs/1708.03384} {arXiv:1708.03384 [astro-ph.HE]}
  \BibitemShut {NoStop}%
\bibitem [{\citenamefont {{J. T. Linnemann for the HAWC
  Collaboration}}(2019)}]{HAWC_LIV_CPT}%
  \BibitemOpen
  \bibfield  {author} {\bibinfo {author} {\bibnamefont {{J. T. Linnemann }}} (\bibinfo {collaboration} {HAWC Collaboration}),\ }in\ \href@noop
  {} {\emph {\bibinfo {booktitle} {{8th Meeting on CPT and Lorentz Symmetry
  (CPT'19) Bloomington, Indiana, USA, May 12-16, 2019}}}}\ (\bibinfo {year}
  {2019})\BibitemShut {NoStop}%
\bibitem [{\citenamefont {Martínez-Huerta}\ \emph {et~al.}(2019)\citenamefont
  {Martínez-Huerta}, \citenamefont {Marinelli}, \citenamefont {Linnemann},\
  and\ \citenamefont {Lundeen}}]{HAWC_ICRC19_Humberto}%
  \BibitemOpen
  \bibfield  {author} {\bibinfo {author} {\bibfnamefont {H.}~\bibnamefont
  {Martínez-Huerta}}, \bibinfo {author} {\bibfnamefont {S.}~\bibnamefont
  {Marinelli}}, \bibinfo {author} {\bibfnamefont {J.~T.}\ \bibnamefont
  {Linnemann}}, \ and\ \bibinfo {author} {\bibfnamefont {J.}~\bibnamefont
  {Lundeen}} (\bibinfo {collaboration} {HAWC Collaboration}),\ }
  \href{https://pos.sissa.it/358/738/}{\bibinfo{journal}{Proc. of Sci. ICRC2019} {\bf \bibinfo{volume}{738}}}\ (\bibinfo {year}
  {2020})\ \Eprint {http://arxiv.org/abs/1908.09614} {arXiv:1908.09614
  [astro-ph.HE]} \BibitemShut {NoStop}%
\bibitem [{\citenamefont {Marinelli}(2019)}]{Sam_Thesis}%
  \BibitemOpen
  \bibfield  {author} {\bibinfo {author} {\bibfnamefont {S.}~\bibnamefont
  {Marinelli}},\ }\href@noop {} {\enquote {\bibinfo {title} {{PhD Thesis,
  Michigan State University}},}\ } (\bibinfo {year} {2019}),\ \bibinfo {note}
  {\url{www.hawc-observatory.org/publications/##thesis}}\BibitemShut {NoStop}%
\bibitem [{\citenamefont {Rubtsov}\ \emph {et~al.}(2017)\citenamefont
  {Rubtsov}, \citenamefont {Satunin},\ and\ \citenamefont
  {Sibiryakov}}]{rubstov_MULTI-TEV}%
  \BibitemOpen
  \bibfield  {author} {\bibinfo {author} {\bibfnamefont {G.}~\bibnamefont
  {Rubtsov}}, \bibinfo {author} {\bibfnamefont {P.}~\bibnamefont {Satunin}}, \
  and\ \bibinfo {author} {\bibfnamefont {S.}~\bibnamefont {Sibiryakov}},\
  }\href {\doibase 10.1088/1475-7516/2017/05/049} {\bibfield  {journal}
  {\bibinfo  {journal} {J. Cosmol. Astropart. Phys.}\ }\textbf {\bibinfo {volume} {05}},\ \bibinfo
  {pages} {049} (\bibinfo {year} {2017})},\ \Eprint
  {http://arxiv.org/abs/1611.10125} {arXiv:1611.10125 [astro-ph.HE]}
  \BibitemShut {NoStop}%
\bibitem [{\citenamefont {Schreck}(2014)}]{Schreck:2013paa}%
  \BibitemOpen
  \bibfield  {author} {\bibinfo {author} {\bibfnamefont {M.}~\bibnamefont
  {Schreck}},\ }in\ \href {\doibase 10.1142/9789814566438_0044} {\emph
  {\bibinfo {booktitle} {{Proceedings, 6th Meeting on CPT and Lorentz Symmetry
  (CPT 13): Bloomington, Indiana, USA, June 17-21, 2013}}}}\ (\bibinfo {year}
  {2014})\ pp.\ \bibinfo {pages} {176--179},\ \Eprint
  {http://arxiv.org/abs/1310.5159} {arXiv:1310.5159 [hep-ph]} \BibitemShut
  {NoStop}%
\bibitem [{\citenamefont {Coleman}\ and\ \citenamefont
  {Glashow}(1999)}]{Coleman:1998ti}%
  \BibitemOpen
  \bibfield  {author} {\bibinfo {author} {\bibfnamefont {S.~R.}\ \bibnamefont
  {Coleman}}\ and\ \bibinfo {author} {\bibfnamefont {S.~L.}\ \bibnamefont
  {Glashow}},\ }\href {\doibase 10.1103/PhysRevD.59.116008} {\bibfield
  {journal} {\bibinfo  {journal} {Phys. Rev.}\ }\textbf {\bibinfo {volume}
  {D59}},\ \bibinfo {pages} {116008} (\bibinfo {year} {1999})},\ \Eprint
  {http://arxiv.org/abs/hep-ph/9812418} {arXiv:hep-ph/9812418 [hep-ph]}
  \BibitemShut {NoStop}%
\bibitem [{\citenamefont {Galaverni}\ and\ \citenamefont
  {Sigl}(2008{\natexlab{a}})}]{GUNTER-PH}%
  \BibitemOpen
  \bibfield  {author} {\bibinfo {author} {\bibfnamefont {M.}~\bibnamefont
  {Galaverni}}\ and\ \bibinfo {author} {\bibfnamefont {G.}~\bibnamefont
  {Sigl}},\ }\href {\doibase 10.1103/PhysRevLett.100.021102} {\bibfield
  {journal} {\bibinfo  {journal} {Phys. Rev. Lett.}\ }\textbf {\bibinfo
  {volume} {100}},\ \bibinfo {pages} {021102} (\bibinfo {year}
  {2008}{\natexlab{a}})},\ \Eprint {http://arxiv.org/abs/0708.1737}
  {arXiv:0708.1737 [astro-ph]} \BibitemShut {NoStop}%
\bibitem [{\citenamefont {Galaverni}\ and\ \citenamefont
  {Sigl}(2008{\natexlab{b}})}]{GUNTER-PD}%
  \BibitemOpen
  \bibfield  {author} {\bibinfo {author} {\bibfnamefont {M.}~\bibnamefont
  {Galaverni}}\ and\ \bibinfo {author} {\bibfnamefont {G.}~\bibnamefont
  {Sigl}},\ }\href {\doibase 10.1103/PhysRevD.78.063003} {\bibfield  {journal}
  {\bibinfo  {journal} {Phys. Rev.}\ }\textbf {\bibinfo {volume} {D78}},\
  \bibinfo {pages} {063003} (\bibinfo {year} {2008}{\natexlab{b}})},\ \Eprint
  {http://arxiv.org/abs/0807.1210} {arXiv:0807.1210 [astro-ph]} \BibitemShut
  {NoStop}%
\bibitem [{\citenamefont {Amelino-Camelia}(2001{\natexlab{b}})}]{DIS1}%
  \BibitemOpen
  \bibfield  {author} {\bibinfo {author} {\bibfnamefont {G.}~\bibnamefont
  {Amelino-Camelia}},\ }\href {\doibase 10.1038/35074035} {\bibfield  {journal}
  {\bibinfo  {journal} {Nature}\ }\textbf {\bibinfo {volume} {410}},\ \bibinfo
  {pages} {1065} (\bibinfo {year} {2001}{\natexlab{b}})},\ \Eprint
  {http://arxiv.org/abs/gr-qc/0104086} {arXiv:gr-qc/0104086 [gr-qc]}
  \BibitemShut {NoStop}%
\bibitem [{\citenamefont {Ahluwalia}(1999)}]{DIS2}%
  \BibitemOpen
  \bibfield  {author} {\bibinfo {author} {\bibfnamefont {D.~V.}\ \bibnamefont
  {Ahluwalia}},\ }\href {\doibase 10.1038/18325} {\bibfield  {journal}
  {\bibinfo  {journal} {Nature}\ }\textbf {\bibinfo {volume} {398}},\ \bibinfo
  {pages} {199} (\bibinfo {year} {1999})},\ \Eprint
  {http://arxiv.org/abs/gr-qc/9903074} {arXiv:gr-qc/9903074 [gr-qc]}
  \BibitemShut {NoStop}%
\bibitem [{\citenamefont {Amelino-Camelia}\ \emph {et~al.}(1998)\citenamefont
  {Amelino-Camelia}, \citenamefont {Ellis}, \citenamefont {Mavromatos},
  \citenamefont {Nanopoulos},\ and\ \citenamefont {Sarkar}}]{DIS3}%
  \BibitemOpen
  \bibfield  {author} {\bibinfo {author} {\bibfnamefont {G.}~\bibnamefont
  {Amelino-Camelia}}, \bibinfo {author} {\bibfnamefont {J.~R.}\ \bibnamefont
  {Ellis}}, \bibinfo {author} {\bibfnamefont {N.~E.}\ \bibnamefont
  {Mavromatos}}, \bibinfo {author} {\bibfnamefont {D.~V.}\ \bibnamefont
  {Nanopoulos}}, \ and\ \bibinfo {author} {\bibfnamefont {S.}~\bibnamefont
  {Sarkar}},\ }\href {\doibase 10.1038/31647} {\bibfield  {journal} {\bibinfo
  {journal} {Nature}\ }\textbf {\bibinfo {volume} {393}},\ \bibinfo {pages}
  {763} (\bibinfo {year} {1998})},\ \Eprint
  {http://arxiv.org/abs/astro-ph/9712103} {arXiv:astro-ph/9712103 [astro-ph]}
  \BibitemShut {NoStop}%
\bibitem [{\citenamefont {Jacobson}\ \emph {et~al.}(2003)\citenamefont
  {Jacobson}, \citenamefont {Liberati},\ and\ \citenamefont
  {Mattingly}}]{JACOB}%
  \BibitemOpen
  \bibfield  {author} {\bibinfo {author} {\bibfnamefont {T.}~\bibnamefont
  {Jacobson}}, \bibinfo {author} {\bibfnamefont {S.}~\bibnamefont {Liberati}},
  \ and\ \bibinfo {author} {\bibfnamefont {D.}~\bibnamefont {Mattingly}},\
  }\href {\doibase 10.1103/PhysRevD.67.124011} {\bibfield  {journal} {\bibinfo
  {journal} {Phys. Rev. D}\ }\textbf {\bibinfo {volume} {67}},\ \bibinfo
  {pages} {124011} (\bibinfo {year} {2003})}\BibitemShut {NoStop}%
\bibitem [{\citenamefont {Scully}\ and\ \citenamefont
  {Stecker}(2009)}]{Stecker2009}%
  \BibitemOpen
  \bibfield  {author} {\bibinfo {author} {\bibfnamefont {S.~T.}\ \bibnamefont
  {Scully}}\ and\ \bibinfo {author} {\bibfnamefont {F.~W.}\ \bibnamefont
  {Stecker}},\ }\href {\doibase 10.1016/j.astropartphys.2009.01.002} {\bibfield
   {journal} {\bibinfo  {journal} {Astropart. Phys.}\ }\textbf {\bibinfo
  {volume} {31}},\ \bibinfo {pages} {220} (\bibinfo {year} {2009})},\ \Eprint
  {http://arxiv.org/abs/0811.2230} {arXiv:0811.2230 [astro-ph]} \BibitemShut
  {NoStop}%
\bibitem [{\citenamefont {Stecker}\ and\ \citenamefont
  {Scully}(2009)}]{Stecker2009NJ}%
  \BibitemOpen
  \bibfield  {author} {\bibinfo {author} {\bibfnamefont {F.~W.}\ \bibnamefont
  {Stecker}}\ and\ \bibinfo {author} {\bibfnamefont {S.~T.}\ \bibnamefont
  {Scully}},\ }\href {\doibase 10.1088/1367-2630/11/8/085003} {\bibfield
  {journal} {\bibinfo  {journal} {New J. Phys.}\ }\textbf {\bibinfo {volume}
  {11}},\ \bibinfo {pages} {085003} (\bibinfo {year} {2009})},\ \Eprint
  {http://arxiv.org/abs/0906.1735} {arXiv:0906.1735 [astro-ph.HE]} \BibitemShut
  {NoStop}%
\bibitem [{\citenamefont {Xu}\ and\ \citenamefont {Ma}(2016)}]{GRB-LIV}%
  \BibitemOpen
  \bibfield  {author} {\bibinfo {author} {\bibfnamefont {H.}~\bibnamefont
  {Xu}}\ and\ \bibinfo {author} {\bibfnamefont {B.-Q.}\ \bibnamefont {Ma}},\
  }\href {\doibase 10.1016/j.physletb.2016.07.044} {\bibfield  {journal}
  {\bibinfo  {journal} {Phys. Lett.}\ }\textbf {\bibinfo {volume} {B760}},\
  \bibinfo {pages} {602} (\bibinfo {year} {2016})},\ \Eprint
  {http://arxiv.org/abs/1607.08043} {arXiv:1607.08043 [hep-ph]} \BibitemShut
  {NoStop}%
\bibitem [{\citenamefont {Zitzer}(2013)}]{VERITAS-LIV}%
  \BibitemOpen
  \bibfield  {author} {\bibinfo {author} {\bibfnamefont {B.}~\bibnamefont
  {Zitzer}} (\bibinfo {collaboration} {VERITAS}),\ }in\ \href@noop {} {\emph
  {\bibinfo {booktitle} {{Proceedings, 33rd International Cosmic Ray Conference
  (ICRC2013): Rio de Janeiro, Brazil, July 2-9, 2013}}}}\ (\bibinfo {year}
  {2013})\ p.\ \bibinfo {pages} {1147},\ \Eprint
  {http://arxiv.org/abs/1307.8382} {arXiv:1307.8382 [astro-ph.HE]} \BibitemShut
  {NoStop}%
\bibitem [{\citenamefont {Otte}(2012)}]{PULSARS-LIV}%
  \BibitemOpen
  \bibfield  {author} {\bibinfo {author} {\bibfnamefont {A.~N.}\ \bibnamefont
  {Otte}},\ }in\ \href {\doibase 10.7529/ICRC2011/V07/1302} {\emph {\bibinfo
  {booktitle} {{Proceedings, 32nd International Cosmic Ray Conference (ICRC
  2011): Beijing, China, August 11-18, 2011}}}},\ Vol.~\bibinfo {volume} {7}\
  (\bibinfo {year} {2012})\ pp.\ \bibinfo {pages} {256--259},\ \Eprint
  {http://arxiv.org/abs/1208.2033} {arXiv:1208.2033 [astro-ph.HE]} \BibitemShut
  {NoStop}%
\bibitem [{\citenamefont {Abramowski}\ \emph {et~al.}(2011)\citenamefont
  {Abramowski} \emph {et~al.}}]{HESS-LIV}%
  \BibitemOpen
  \bibfield  {author} {\bibinfo {author} {\bibfnamefont {A.}~\bibnamefont
  {Abramowski}} \emph {et~al.} (\bibinfo {collaboration} {H.E.S.S.}),\ }\href
  {\doibase 10.1016/j.astropartphys.2011.01.007} {\bibfield  {journal}
  {\bibinfo  {journal} {Astropart. Phys.}\ }\textbf {\bibinfo {volume} {34}},\
  \bibinfo {pages} {738} (\bibinfo {year} {2011})},\ \Eprint
  {http://arxiv.org/abs/1101.3650} {arXiv:1101.3650 [astro-ph.HE]} \BibitemShut
  {NoStop}%
\bibitem [{\citenamefont {Lang}\ \emph {et~al.}(2019)\citenamefont {Lang},
  \citenamefont {Martínez-Huerta},\ and\ \citenamefont
  {de~Souza}}]{Lang:2018yog}%
  \BibitemOpen
  \bibfield  {author} {\bibinfo {author} {\bibfnamefont {R.~G.}\ \bibnamefont
  {Lang}}, \bibinfo {author} {\bibfnamefont {H.}~\bibnamefont
  {Martínez-Huerta}}, \ and\ \bibinfo {author} {\bibfnamefont
  {V.}~\bibnamefont {de~Souza}},\ }\href {\doibase 10.1103/PhysRevD.99.043015}
  {\bibfield  {journal} {\bibinfo  {journal} {Phys. Rev.}\ }\textbf {\bibinfo
  {volume} {D99}},\ \bibinfo {pages} {043015} (\bibinfo {year} {2019})},\
  \Eprint {http://arxiv.org/abs/1810.13215} {arXiv:1810.13215 [astro-ph.HE]}
  \BibitemShut {NoStop}%
\bibitem [{\citenamefont {Ellis}\ \emph {et~al.}(2006)\citenamefont {Ellis},
  \citenamefont {Mavromatos}, \citenamefont {Nanopoulos}, \citenamefont
  {Sakharov},\ and\ \citenamefont {Sarkisyan}}]{E1}%
  \BibitemOpen
  \bibfield  {author} {\bibinfo {author} {\bibfnamefont {J.~R.}\ \bibnamefont
  {Ellis}}, \bibinfo {author} {\bibfnamefont {N.~E.}\ \bibnamefont
  {Mavromatos}}, \bibinfo {author} {\bibfnamefont {D.~V.}\ \bibnamefont
  {Nanopoulos}}, \bibinfo {author} {\bibfnamefont {A.~S.}\ \bibnamefont
  {Sakharov}}, \ and\ \bibinfo {author} {\bibfnamefont {E.~K.~G.}\ \bibnamefont
  {Sarkisyan}},\ }\href {\doibase 10.1016/j.astropartphys.2006.04.001,
  10.1016/j.astropartphys.2007.12.003} {\bibfield  {journal} {\bibinfo
  {journal} {Astropart. Phys.}\ }\textbf {\bibinfo {volume} {25}},\ \bibinfo
  {pages} {402} (\bibinfo {year} {2006})},\ \bibinfo {note} {[Erratum:
  Astropart. Phys.29,158(2008)]},\ \Eprint {http://arxiv.org/abs/0712.2781}
  {arXiv:0712.2781 [astro-ph]} \BibitemShut {NoStop}%
\bibitem [{\citenamefont {Albert}\ \emph {et~al.}(2008)\citenamefont {Albert}
  \emph {et~al.}}]{E2}%
  \BibitemOpen
  \bibfield  {author} {\bibinfo {author} {\bibfnamefont {J.}~\bibnamefont
  {Albert}} \emph {et~al.} (\bibinfo {collaboration} {MAGIC, Other
  Contributors}),\ }\href {\doibase 10.1016/j.physletb.2008.08.053} {\bibfield
  {journal} {\bibinfo  {journal} {Phys. Lett.}\ }\textbf {\bibinfo {volume}
  {B668}},\ \bibinfo {pages} {253} (\bibinfo {year} {2008})},\ \Eprint
  {http://arxiv.org/abs/0708.2889} {arXiv:0708.2889 [astro-ph]} \BibitemShut
  {NoStop}%
\bibitem [{\citenamefont {Ellis}\ \emph {et~al.}(2003)\citenamefont {Ellis},
  \citenamefont {Mavromatos}, \citenamefont {Nanopoulos},\ and\ \citenamefont
  {Sakharov}}]{E3}%
  \BibitemOpen
  \bibfield  {author} {\bibinfo {author} {\bibfnamefont {J.~R.}\ \bibnamefont
  {Ellis}}, \bibinfo {author} {\bibfnamefont {N.~E.}\ \bibnamefont
  {Mavromatos}}, \bibinfo {author} {\bibfnamefont {D.~V.}\ \bibnamefont
  {Nanopoulos}}, \ and\ \bibinfo {author} {\bibfnamefont {A.~S.}\ \bibnamefont
  {Sakharov}},\ }\href {\doibase 10.1051/0004-6361:20030263} {\bibfield
  {journal} {\bibinfo  {journal} {Astron. Astrophys.}\ }\textbf {\bibinfo
  {volume} {402}},\ \bibinfo {pages} {409} (\bibinfo {year}
  {2003})}\BibitemShut {NoStop}%
\bibitem [{\citenamefont {Schreck}(2017)}]{VCR-17}%
  \BibitemOpen
  \bibfield  {author} {\bibinfo {author} {\bibfnamefont {M.}~\bibnamefont
  {Schreck}},\ }\href {\doibase 10.1103/PhysRevD.96.095026} {\bibfield
  {journal} {\bibinfo  {journal} {Phys. Rev.}\ }\textbf {\bibinfo {volume}
  {D96}},\ \bibinfo {pages} {095026} (\bibinfo {year} {2017})},\ \Eprint
  {http://arxiv.org/abs/1702.03171} {arXiv:1702.03171 [hep-ph]} \BibitemShut
  {NoStop}%
\bibitem [{\citenamefont {Mart\'inez-Huerta}\ and\ \citenamefont
  {P\'erez-Lorenzana}(2016)}]{Proc2}%
  \BibitemOpen
  \bibfield  {author} {\bibinfo {author} {\bibfnamefont {H.}~\bibnamefont
  {Mart\'inez-Huerta}}\ and\ \bibinfo {author} {\bibfnamefont {A.}~\bibnamefont
  {P\'erez-Lorenzana}},\ } \href {\doibase
  10.1088/1742-6596/761/1/012035} {\bibfield  {journal} {\bibinfo  {journal}
  {J. Phys. Conf. Ser.}\ }\textbf {\bibinfo {volume} {761}},\ \bibinfo {pages}
  {012035} (\bibinfo {year} {2016})},\ \Eprint
  {http://arxiv.org/abs/1609.07185} {arXiv:1609.07185 [astro-ph.HE]}
  \BibitemShut {NoStop}%
\bibitem [{\citenamefont {Martínez-Huerta}\ and\ \citenamefont
  {Pérez-Lorenzana}(2017)}]{Martinez-Huerta:2017ntv}%
  \BibitemOpen
  \bibfield  {author} {\bibinfo {author} {\bibfnamefont {H.}~\bibnamefont
  {Martínez-Huerta}}\ and\ \bibinfo {author} {\bibfnamefont {A.}~\bibnamefont
  {Pérez-Lorenzana}},\ }\bibfield  {booktitle} \href {\doibase
  10.1088/1742-6596/866/1/012006} {\bibfield  {journal} {\bibinfo  {journal}
  {J. Phys. Conf. Ser.}\ }\textbf {\bibinfo {volume} {866}},\ \bibinfo {pages}
  {012006} (\bibinfo {year} {2017})},\ \Eprint
  {http://arxiv.org/abs/1702.00913} {arXiv:1702.00913 [hep-ph]} \BibitemShut
  {NoStop}%
\bibitem [{\citenamefont {Malone}(2019)}]{HAWC_ICRC19_Kelly}%
  \BibitemOpen
  \bibfield  {author} {\bibinfo {author} {\bibfnamefont {K.}~\bibnamefont
  {Malone}} (\bibinfo {collaboration} {HAWC Collaboration}),\ } 
  \href{https://pos.sissa.it/358/734/}{\bibinfo{journal}{Proc. of Sci. ICRC2019} {\bf \bibinfo{volume}{734}}}\ (\bibinfo {year}
  {2020}) \ \Eprint {http://arxiv.org/abs/1908.07059} {arXiv:1908.07059
  [astro-ph.HE]} \BibitemShut {NoStop}%
\bibitem [{Sup(2019)}]{Supplemental}%
  \BibitemOpen
  {\bibinfo {title} {{See the Supplemental Material section for details of the photon energy limit calculation and a more comprehensive list of the LIV limits, including its corresponding values in the SME framework}}}\ \BibitemShut {NoStop}%
\bibitem [{\citenamefont {Chernoff}(1954)}]{chernoff1954}%
  \BibitemOpen
  \bibfield  {author} {\bibinfo {author} {\bibfnamefont {H.}~\bibnamefont
  {Chernoff}},\ }\href {\doibase 10.1214/aoms/1177728725} {\bibfield  {journal}
  {\bibinfo  {journal} {Ann. Math. Stat.}\ }\textbf {\bibinfo {volume}
  {25}},\ \bibinfo {pages} {573} (\bibinfo {year} {1954})}\BibitemShut
  {NoStop}%
\bibitem [{\citenamefont {Manchester}\ \emph {et~al.}(2005)\citenamefont
  {Manchester}, \citenamefont {Hobbs}, \citenamefont {Teoh},\ and\
  \citenamefont {Hobbs}}]{ATNF_Catalog}%
  \BibitemOpen
  \bibfield  {author} {\bibinfo {author} {\bibfnamefont {R.~N.}\ \bibnamefont
  {Manchester}}, \bibinfo {author} {\bibfnamefont {G.~B.}\ \bibnamefont
  {Hobbs}}, \bibinfo {author} {\bibfnamefont {A.}~\bibnamefont {Teoh}}, \ and\
  \bibinfo {author} {\bibfnamefont {M.}~\bibnamefont {Hobbs}},\ }\href
  {\doibase 10.1086/428488} {\bibfield  {journal} {\bibinfo  {journal} {Astron.
  J.}\ }\textbf {\bibinfo {volume} {129}},\ \bibinfo {pages} {1993} (\bibinfo
  {year} {2005})},\ \Eprint {http://arxiv.org/abs/astro-ph/0412641}
  {arXiv:astro-ph/0412641 [astro-ph]} \BibitemShut {NoStop}%
\bibitem [{\citenamefont {Kosteleck\'y}\ and\ \citenamefont
  {Mewes}(2009)}]{kos09}%
  \BibitemOpen
  \bibfield  {author} {\bibinfo {author} {\bibfnamefont {V.~A.}\ \bibnamefont
  {Kosteleck\'y}}\ and\ \bibinfo {author} {\bibfnamefont {M.}~\bibnamefont
  {Mewes}},\ }\href {\doibase 10.1103/PhysRevD.80.015020} {\bibfield  {journal}
  {\bibinfo  {journal} {Phys. Rev. D}\ }\textbf {\bibinfo {volume} {80}},\
  \bibinfo {pages} {015020} (\bibinfo {year} {2009})}\BibitemShut {NoStop}%
\bibitem [{\citenamefont {Kostelecky}\ and\ \citenamefont
  {Russell}(2011)}]{SME_tables}%
  \BibitemOpen
  \bibfield  {author} {\bibinfo {author} {\bibfnamefont {V.~A.}\ \bibnamefont
  {Kostelecky}}\ and\ \bibinfo {author} {\bibfnamefont {N.}~\bibnamefont
  {Russell}},\ }\href {\doibase 10.1103/RevModPhys.83.11} {\bibfield  {journal}
  {\bibinfo  {journal} {Rev. Mod. Phys.}\ }\textbf {\bibinfo {volume} {83}},\
  \bibinfo {pages} {11} (\bibinfo {year} {2011})},\ \Eprint
  {http://arxiv.org/abs/0801.0287} {arXiv:0801.0287 [hep-ph]} \BibitemShut
  {NoStop}%
\bibitem [{\citenamefont {De~Angelis}\ \emph {et~al.}(2013)\citenamefont
  {De~Angelis}, \citenamefont {Galanti},\ and\ \citenamefont
  {Roncadelli}}]{Angelis}%
  \BibitemOpen
  \bibfield  {author} {\bibinfo {author} {\bibfnamefont {A.}~\bibnamefont
  {De~Angelis}}, \bibinfo {author} {\bibfnamefont {G.}~\bibnamefont {Galanti}},
  \ and\ \bibinfo {author} {\bibfnamefont {M.}~\bibnamefont {Roncadelli}},\
  }\href {\doibase 10.1093/mnras/stt684} {\bibfield  {journal} {\bibinfo
  {journal} {Mon. Not. R. Astron. Soc.}\ }\textbf {\bibinfo {volume} {432}},\
  \bibinfo {pages} {3245} (\bibinfo {year} {2013})},\ \Eprint
  {http://arxiv.org/abs/1302.6460} {arXiv:1302.6460 [astro-ph.HE]} \BibitemShut
  {NoStop}%
\bibitem [{\citenamefont {Gould}\ and\ \citenamefont
  {Schréder}(1966)}]{Gould:1966pza}%
  \BibitemOpen
  \bibfield  {author} {\bibinfo {author} {\bibfnamefont {R.}~\bibnamefont
  {Gould}}\ and\ \bibinfo {author} {\bibfnamefont {G.}~\bibnamefont
  {Schréder}},\ }\href {\doibase 10.1103/PhysRevLett.16.252} {\bibfield
  {journal} {\bibinfo  {journal} {Phys. Rev. Lett.}\ }\textbf {\bibinfo
  {volume} {16}},\ \bibinfo {pages} {252} (\bibinfo {year} {1966})}\BibitemShut
  {NoStop}%
\bibitem [{\citenamefont {Gould}\ and\ \citenamefont
  {Schreder}(1967{\natexlab{a}})}]{Gould:1967zza}%
  \BibitemOpen
  \bibfield  {author} {\bibinfo {author} {\bibfnamefont {R.~J.}\ \bibnamefont
  {Gould}}\ and\ \bibinfo {author} {\bibfnamefont {G.~P.}\ \bibnamefont
  {Schreder}},\ }\href {\doibase 10.1103/PhysRev.155.1408} {\bibfield
  {journal} {\bibinfo  {journal} {Phys. Rev.}\ }\textbf {\bibinfo {volume}
  {155}},\ \bibinfo {pages} {1408} (\bibinfo {year}
  {1967}{\natexlab{a}})}\BibitemShut {NoStop}%
\bibitem [{\citenamefont {Gould}\ and\ \citenamefont
  {Schreder}(1967{\natexlab{b}})}]{Gould:1967zzb}%
  \BibitemOpen
  \bibfield  {author} {\bibinfo {author} {\bibfnamefont {R.~J.}\ \bibnamefont
  {Gould}}\ and\ \bibinfo {author} {\bibfnamefont {G.~P.}\ \bibnamefont
  {Schreder}},\ }\href {\doibase 10.1103/PhysRev.155.1404} {\bibfield
  {journal} {\bibinfo  {journal} {Phys. Rev.}\ }\textbf {\bibinfo {volume}
  {155}},\ \bibinfo {pages} {1404} (\bibinfo {year}
  {1967}{\natexlab{b}})}\BibitemShut {NoStop}%
\end{thebibliography}
\end{document}